\documentclass{article}
\usepackage{arxiv}            
\usepackage{amsmath, amssymb}
\usepackage{natbib}           
\usepackage{hyperref}
\usepackage{geometry}
\usepackage{algorithm, algorithmic}
\usepackage{graphicx}
\usepackage{subfigure}
\usepackage{booktabs} 
\usepackage[table]{xcolor}        
\definecolor{noiseLow}{gray}{0.96}   
\definecolor{noiseMid}{gray}{0.90}   
\definecolor{noiseHi} {gray}{0.80}   

\newcommand{\lownoise} {\rowcolor{noiseLow}}
\newcommand{\midnoise} {\rowcolor{noiseMid}}
\newcommand{\highnoise}{\rowcolor{noiseHi}}

\newcommand*{\blauw}[1]{\textcolor{blue}{#1}}

\usepackage{hyperref}

\usepackage{amsmath}
\usepackage{amssymb}
\usepackage{mathtools}
\usepackage{amsthm}

\usepackage[capitalize,noabbrev]{cleveref}

\theoremstyle{plain}

\theoremstyle{definition}

\theoremstyle{remark}

\usepackage[textsize=tiny]{todonotes}

\begin{document}





\title{FlowBack-Adjoint: Physics-Aware and Energy-Guided\\ Conditional Flow-Matching for All-Atom Protein Backmapping}
\author{
  Alex Berlaga\\
  Department of Chemistry\\
  University of Chicago, Chicago, IL 60637, USA\\
  \texttt{berlaga@uchicago.edu}
  \And
  Michael S. Jones\\
  Lawrence Livermore National Laboratory, 
  Livermore, CA 94550, USA\\
  \texttt{jones313@llnl.gov}
  \And
  Andrew L.\ Ferguson\\
  Pritzker School of Molecular Engineering and Department of Chemistry\\
  University of Chicago, Chicago, IL 60637, USA\\
  \texttt{andrewferguson@uchicago.edu}
}

\date{\hspace{0.1in}}

\maketitle






\begin{abstract}
Coarse-grained (CG) molecular models of proteins can substantially increase the time and length scales accessible to molecular dynamics simulations of proteins, but recovery of accurate all-atom (AA) ensembles from CG simulation trajectories can be essential for exposing molecular mechanisms of folding and docking and for calculation of physical properties requiring atomistic detail. The recently reported deep generative model \textsc{FlowBack} restores AA detail to protein C$_\alpha$ traces using a flow-matching architecture and demonstrates state-of-the-art performance in generation of AA structural ensembles. Training, however, is performed exclusively on structural data and the absence of any awareness of interatomic energies or forces within training results in small fractions of incorrect bond lengths, atomic clashes, and otherwise high-energy structures. In this work, we introduce \textsc{FlowBack-Adjoint} as a lightweight enhancement that upgrades the pre-trained \textsc{FlowBack} model through a one-time, physics-aware post-training pass. Auxiliary contributions to the flow introduce physical awareness of bond lengths and Lennard-Jones interactions and gradients of a molecular mechanics force field energy are incorporated via adjoint matching to steer the \textsc{FlowBack-Adjoint} vector field to produce lower-energy configurations. In benchmark tests against \textsc{FlowBack}, \textsc{FlowBack-Adjoint} lowers single-point energies by a median of $\sim$78 kcal/mol.residue, reduces errors in bond lengths by $>$92\%, eliminates $>$98\% of molecular clashes, maintains excellent diversity of the AA configurational ensemble, and produces configurations capable of initializing stable all-atom molecular dynamics simulations without requiring energy relaxation. We propose \textsc{FlowBack-Adjoint} as an accurate and efficient physics-aware deep generative model for AA backmapping from C$_\alpha$ traces from structure prediction tools, coarse-grained simulations, or low-resolution experimental data for downstream atomistic modeling.
\end{abstract}



\section{Introduction} 

The last five years have witnessed a paradigm shift in computational structural biology that has transformed protein structure prediction from a grand challenge into a routine operation. Breakthrough deep-learning frameworks including \textsc{AlphaFold2} \citep{jumper2021highly}, \textsc{RoseTTAFold} \citep{rosettafold}, \textsc{ESMFold} \citep{esmfold}, \textsc{OmegaFold} \citep{Wu2022}, and \textsc{RFDiffusion} \citep{rfdiffusion} possess backbone prediction accuracies at near-experimental resolution, and \textsc{AlphaFold2} alone has populated the public AlphaFold Protein Structure Database with $\gtrsim$ 250 million high-confidence models. In reality, proteins exhibit conformational ensembles rather than a single rigid conformation \citep{wei2016, wolf2025}. Side-chain flexibility, in particular, is an important contributor to conformational entropy that impacts folding, stability, and molecular interactions \citep{Cagiada2025, Berka2010, Bachmann2011}. Powerful backbone predictors such as \textsc{AlphaFold2} produce deterministic side-chain placements, omitting the conformational diversity and breadth of side-chain rotameric states or alternate conformers. \textsc{AlphaFold3} has replaced this deterministic stage with a generative diffusion head \citep{Alphafold3}, but offers no explicit mechanism to ensure that its samples generate a physically plausible and low-energy structural ensemble. This limitation hinders downstream tasks such as protein or ligand docking \citep{Xu2022, Yong2008}, which often require a diverse set of low-energy all-atom structures covering the thermally-relevant configurational space for accurate predictions.

The conventional approach to generate protein conformational ensembles is to perform all-atom molecular dynamics (MD) simulations. These calculations, however, become exceedingly expensive for large proteins and are generally restricted to time scales of $\mu$s-ms \citep{Buch2010}. Accordingly, several generative methods have recently emerged to efficiently emulate protein ensemble sampling \citep{wolf2025}. For example, \textsc{AlphaFlow} fine-tunes the \textsc{AlphaFold2} model on MD ensembles to predict multiple conformations \citep{jing2024alphafold, jumper2021highly} and diffusion-based approaches like \textsc{ConfDiff} introduce stochasticity into structure prediction \citep{wang2024protein}. These strategies can generate realistic backbone ensembles, but often still yield only a single most-probable side-chain configuration per backbone. Meanwhile, other tools focus on backbone dynamics. For example, \textsc{BBFlow} uses flow-matching to sample protein backbone conformations conditioned on a given structure \citep{wolf2025} and \textsc{BioEmu} predicts equilibrium structural ensembles using a multiple sequence alignment-based diffusion architecture \citep{Lewis2024}. Such models capture protein flexibility at the backbone level, yet they do not explicitly address the rich combinatorial space of side-chain conformations. Computational side-chain packing methods such as \textsc{Rosetta} \citep{rosetta} or \textsc{SCWRL} \citep{scwrl} and recent machine learning-based packing methods such as \textsc{FlowPacker} \citep{Lee2024}, \textsc{AttnPacker} \citep{mcparlton2023}, \textsc{DLPacker} \citep{Misiura2022}, \textsc{HPacker} \citep{visani2023}, and \textsc{PIPPack} \citep{Randolph2023} are typically capable of engaging this diversity challenge but can perform poorly when comparing energies against an ensemble of MD-derived protein configurations. 

In this work, we present a generative modeling tool to fill this gap by transforming backbone-only structures into all-atom ensembles with diverse, physically-plausible, low-energy conformations. Our approach is based on a pre-trained flow-based generative model, \textsc{FlowBack}, originally developed for coarse-grained backmapping \citep{jones2025flowback}. Coarse-grained (CG) models replace groups of atoms with beads comprising groups of atoms thereby reducing the number of degrees of freedom. This resolution reduction enables simulations to be conducted at a fraction of the cost of all-atom MD and the ability to access vastly longer time and length scales \citep{noid2008multiscale,clementi2008coarse,jin2022bottom}. Backmapping can be conceived as the inverse of coarse-graining, \textit{viz.}\ the re-introduction of all-atom (AA) detail to a CG scaffold -- be it a heavy-atom backbone, a bead model, a sparse electron-density trace, or a C$_\alpha$ trace -- by positioning the remaining heavy atoms and side chains so that the resulting structure is chemically consistent with the underlying coarse geometry \citep{wassenaar2014going,lombardi2016cg2aa}. Coarse-graining is a many-to-one operation entailing a loss of resolution wherein multiple AA configurations are consistent with a single CG configuration. Backmapping, therefore, is a one-to-many operation, and it is typically desirable to generate a physically-plausible ensemble of AA configurations -- ideally weighted according to some distribution such as the Boltzmann ensemble -- consistent with a particular CG configuration \citep{wang2022generative,jones2023diamondback,jones2025flowback}. A number of data-driven protein backmapping approaches have been developed in recent years that have explored a variety of machine learning architectures -- including generative-adversarial networks, autoencoders, denoising diffusion probability models, and flow-matching -- and a range of CG resolutions \citep{stieffenhofer2020adversarial,stieffenhofer2021adversarial,stieffenhofer2022benchmarking,li2020backmapping,an2020machine,wang2022generative,shmilovich2022temporally,liu2023backdiff,li2024towards,pang2024simple,heo2024one,angioletti2024herobm,waltmann2025msback,jones2023diamondback,jones2025flowback}. \textsc{FlowBack} stands out as a deep generative approach based on a conditional flow-matching objective that exhibits state-of-the-art performance in the efficient generation of AA structural ensembles compatible with C$_\alpha$ traces that possess fewer clashes and higher diversities of physically-plausible AA conformations compared with other approaches \citep{jones2025flowback}. It is a generic model that works for any protein, scales well to extremely large proteins of lengths far outside the training data, admits straightforward and lightweight fine-tuning for particular proteins or protein families, has been extended to perform simultaneous backmapping of protein-DNA complexes, and is available as an open source Python package \citep{jones2025flowback}.

\textsc{FlowBack} delivers diverse ensembles of physically-reasonable AA protein structures by placing the backbone and side chain atoms in every residue on top of their parent C$_\alpha$ and learning a flow to correctly guide each atom to a distribution of plausible residue conformations. The model is trained on pairs of CG C$_\alpha$ traces and corresponding AA structures curated by coarse-graining AA conformations curated from the Protein Data Bank \citep{10.1093/nar/28.1.235}. A deficiency of this structure-only training objective is that the model possesses no awareness of bonded and non-bonded energetics. The sensitive dependence of the energy upon the structure mean that structurally-plausible configurations may nonetheless be quite high in energy. For example, a 10\% perturbation in a single bond length away from its equilibrium minimum can, for some of the stiffest bonds in a typical protein, lead to a $\sim$13 kcal/mol elevation in the molecular energy. The presence of high-energy configurations in the generated AA structural ensemble means that the model does not produce a distribution consistent with the Boltzmann ensemble at a particular temperature, has no awareness of how to relax conformations in configurational space towards low-energy states by following force vectors on the atoms, and can produce configurations lying tens of kcal/mol.residue above the local minimum in the potential energy landscape. Indeed, some generated configurations can lie so high in energy as to destabilize the integrator if used to directly initialize a MD run without first performing energy relaxation. From a practical perspective, it is possible to subject the \textsc{FlowBack}-generated AA structures to molecular mechanics relaxation, but this can degrade the diversity of configurations by shunting them towards the bottom of local energy minima. From a principled perspective, it would be desirable for the model to inherently generate low-energy ensembles of AA configurations. 

\begin{figure}[h]
    \centering
    \includegraphics[width=\linewidth]{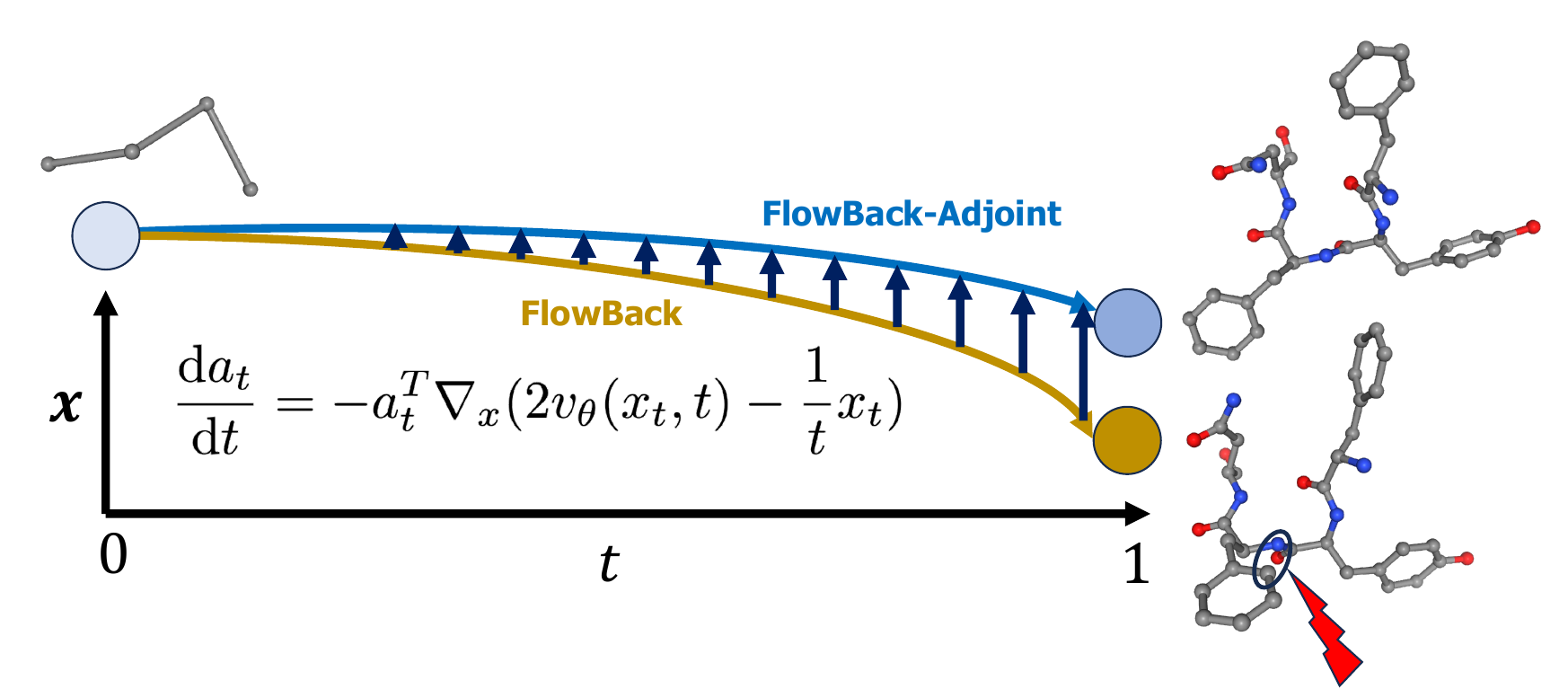}
    \caption{\textsc{FlowBack-Adjoint} improves on the state-of-the art \textsc{FlowBack} \citep{jones2025flowback} conditional flow-matching backmapping model to restore all-atom detail to coarse-grained C$_\alpha$ representations using adjoint matching to incorporate inductive biases from a molecular force field and promote generation of conformationally-diverse and low-energy ensembles of all-atom backmapped structures. The incorporation of physics-aware flows reduces the presence of high-energy configurations in the generated ensemble, such as atom-atom clashes indicated by the red lightning bolt. The lean adjoint ordinary differential equation emerging from the adjoint matching formalism fine-tunes the flow towards low-energy structures.}
    \label{fig:summary}
\end{figure}

The principal contribution of this work is to introduce \textsc{FlowBack-Adjoint} as an enhancement of the pre-trained \textsc{FlowBack} model that updates the conditional flow-matching vector field to steer generation towards low-energy configurations via adjoint matching \citep{domingoenrich2025adjointmatching} (Figure~\ref{fig:summary}). Specifically, after \textsc{FlowBack} has learned to map CG inputs \(x_{0}\) to AA outputs \(x_{1}\) via conditional flow-matching,
we harvest so-called memoryless trajectories by pairing this drift with carefully selected noise schedules that ensure outputs are sampled from a desired distribution and remain independent of randomly initialized inputs. 
We incorporate information on energy and forces via an inductive bias wherein at $t=1$ we evaluate a force field energy \(E(x)\) and back-propagate its gradient \(\nabla_{x}E\) through the time-reversed system. The resulting parameter updates, computed with standard automatic-differentiation tools, nudge the flow toward trajectories that terminate in lower-energy conformations without compromising the structural backmapping and diversity of AA configurations learned by \textsc{FlowBack} or risking instabilities or catastrophic forgetting that may be introduced by a more invasive and direct modification of the conditional flow-matching velocity. Since the energy correction is applied only during fine-tuning, inference speed is unchanged and the model still requires nothing more than the CG trace supplied to the original \textsc{FlowBack}. We incorporate further physics-based inductive biases by gradually activating bond-length constraints and Lennard-Jones interactions during later stages of the flow to promote adherence to these physical constraints.

In benchmark tests against \textsc{FlowBack}, we find that \textsc{FlowBack-Adjoint} lowers single-point energies by a median of $\sim$78 kcal/mol.residue, reduces errors in bond lengths by $>$92\%, eliminates $>$98\% of molecular clashes, and maintains excellent diversity of the AA configurational ensemble. Crucially, \textsc{FlowBack-Adjoint} requires as its only input a C$_\alpha$ trace with which to condition the flow, and is agnostic to its source, permitting compatibility with CG protein simulations using a coarse-grained force field such as MARTINI \citep{souza2021martini}, AWSEM \citep{davtyan2012awsem}, or CALVADOS \citep{tesei2023improved}, incomplete models from low-resolution nuclear magnetic resonance (NMR) or cryo-electron microscopy (cryo-EM) measurements, or structural predictions from computational tools such as \textsc{AlphaFlow} \citep{jing2024alphafold}, \textsc{BBFlow} \citep{wolf2025}, or \textsc{BioEmu} \citep{Lewis2024}. 

Recently, several models have been developed that perform atomistic-resolution protein generation using a variety of machine learning strategies. \textsc{La-Proteina} trains a continuous-time flow-matching model on a hybrid representation in which the backbone is treated explicitly while side-chain identity and geometry live in a latent space, allowing joint sampling of sequence and all-atom structure \citep{geffner2025laproteina}.  \textsc{P(all-atom)} uses coupled diffusion processes that simultaneously emit residue-level tokens and atomic coordinates, so sequence and structure are created coherently in a single forward pass \citep{Qu2024}. Most similar to our approach, \textsc{Energy-Based Alignment} fine-tunes a pretrained denoising-diffusion generator by pairwise matching generated samples' log-probability differences to the differences in their energy returned by a classical force field \citep{lu2025eba}. Although these studies tackle the harder task of full-backbone generation, their evaluations center on matching geometric or statistical properties of existing datasets. In this work, we subject \textsc{FlowBack-Adjoint} to stringent physics-based tests including bond-length fidelity, steric clash counts, configurational diversity, molecular mechanics force field energy assessment, and stability of generated configurations in launching MD simulations without any additional energy relaxation.

\section{Methods} \label{sec:meth}

\subsection{\textsc{FlowBack}}

\textsc{FlowBack} is a deep generative approach based on a conditional flow-matching objective to produce AA structural ensembles of proteins from C$_\alpha$ CG traces \citep{jones2025flowback}. 
It learns a rotationally and translationally equivariant vector field \(v_{\gamma}(x_t,t)\) whose deterministic integration carries a noisy, CG-conditioned prior distribution \(q_{0}\bigl(x\,|\,\mathrm{C}_\alpha\bigr)\) at \(t=0\) to a learned all-atom distribution \(q_{1}(x)\) at \(t=1\). Each generated conformation comes equipped with an exact likelihood and exactly preserves the supplied C$_\alpha$ trace. The prior distribution is constructed by placing every heavy atom in an isotropic Gaussian of variance \(\sigma_{p}^{2}\) around its parent C\(_\alpha\) atom. The noise amplitude \(\sigma_{p}\) is the model's single exposed hyper-parameter that permits users trade off between high structural accuracy (smaller \(\sigma_{p}\)) and diverse configurational ensembles (larger \(\sigma_{p}\)). \textsc{FlowBack} operates without explicit consideration of hydrogen atoms, leaving these to be placed by downstream tools after the learned flow has positioned all of the heavy atoms.

Training of the learned vector field vector field \(v_{\gamma}(x_t,t)\) proceeds entirely without molecular dynamics or structure relaxation, relying instead on all-atom structural databases to define a supervised learning objective. The model was trained over a corpus of 65,360 all-atom protein structures containing 20--1000 residues that were derived from the SidechainNet \citep{king2021sidechainnet} extension of ProteinNet \citep{alquraishi2019proteinnet}, which was itself curated from the Protein Data Bank \citep{10.1093/nar/28.1.235}. Hydrogen atoms were stripped to define the AA targets \(x_{1} \sim q_1(x)\) and the C$_\alpha$ traces extracted to define the corresponding CG conditionings to generate the initial samples \(x_{0}\) from which we learn the flow. For each training batch, a noisy initial structure \(x_{0} \sim q_0(x) = \mathcal{N}\!\bigl(x_0[M],\sigma_{p}^{2}I\bigr)\) is drawn, a random time \(t\in[0,1]\) is chosen, a linearly interpolated mean \(\mu_{t}=tx_{1}+(1-t)x_{0}\) is formed, and i.i.d.\ Gaussian noise of variance \(\sigma_{int}^{2}\) added to the mean to obtain \(x_t\sim\mathcal{N}\!\bigl(\mu_{t},\sigma_{int}^{2}I\bigr)\). The coordinates of the C$_\alpha$ atoms are placed under a mask $M$ to ensure that their coordinates in the final backmapped structure exactly match those in the conditioning C$_\alpha$ trace. However, later testing revealed that the C$_\alpha$ mask had little influence on the final all-atom coordinates, so in retraining \textsc{FlowBack} as a base model for the development of \textsc{FlowBack-Adjoint},  we eliminated the mask in learning the flow. A six-layer equivariant graph neural network (EGNN) is trained to learn the vector field \(v_{\gamma}(x_t,t)\) by regressing against the reference field \(u_{t}=\left(x_{1}-x_{0}\right)\) under an $L1$ loss. Operationally, we ``trick'' the EGNN into learning the drift $v_{\gamma}(x_t,t)$ by training it to predict a one-step-ahead configuration $x_t^\prime = EGNN_\gamma(x_t,t)$ and then computing $v_{\gamma}(x_t,t) = (x_t^\prime - x_t)$.

At inference time, initial structures $x_0$ are generated by initializing the heavy atoms around their parent C\(_\alpha\) atom by drawing from \(x_{0}\sim\mathcal{N}\!\bigl(x_0,\sigma_{p}^{2}I\bigr)\), where the $\sigma_p$ used in inference need not be the same $\sigma_p$ applied in training. The model then uses the learned flow to integrate the ordinary differential equation \(\dot{x}_{t}=v_{\gamma}(x_t,t)\) for 100 Euler steps. 

Amino acids -- with the exception of glycine -- are chiral, meaning their side chains can exist in two non-equivalent mirror-image forms known as L-form and D-form. Natural proteins employ only the L-form, with 99.99\% of the \textsc{FlowBack} training data belongs to this stereoisomeric class. However, the \textsc{FlowBack} EGNN does not explicitly encode chirality constraints and, as a result, without any corrections to the learned flow, a small fraction of generated side chains -- about 3.7\% in the original experiments -- emerge in the incorrect D-form. \textsc{FlowBack} corrects this by detecting side chains that initially appear as D-isomers both early in the trajectory at \(t=0.2\) and at the end at $t = 1$ to, if necessary, perform a corrective mirror reflection. This operation eliminates all chirality errors while slightly raising the incidence of structural clashes.

\subsection{\textsc{FlowBack+LJ/Bonds}} \label{subsec:ljbonds}

\textsc{FlowBack-Adjoint} is a sophistication of \textsc{FlowBack} that employs adjoint matching to fine-tune the learned flow by steering it towards low-energy configurations. As an intermediate stepping stone to this model, we introduce \textsc{FlowBack+LJ/Bonds} as a physically motivated augmentation to \textsc{FlowBack} that adds three auxiliary time-gated velocity fields to the learned drift \(v_{\gamma}\) without changing the network parameters, numerical integrator, or model architecture. \textit{First}, in place of \textsc{FlowBack}'s original discrete side-chain flipping protocol, we inject a continuous, differentiable velocity field \(v_{\text{chiral}}\) for \(t \ge 0.25\) that smoothly steers any D-form residue toward the correct L-form configuration by pushing it along the axis perpendicular to the reflection plane, thereby enforcing native stereochemistry without breaking gradient flow. \textit{Second}, a repulsive Lennard-Jones (LJ) velocity is introduced for \(t \ge 0.85\), wherein heavy-atom pairs separated by more than two covalent bonds and residing within a distance $d$ $<$ 0.42 nm are subjected to a Lennard-Jones interaction under the CHARMM27 force field \citep{CHARMM}. \textit{Third}, for (\(t \ge 0.95\)), a harmonic bond length interaction is added around the equilibrium bond length mandated by the CHARMM27 force field \citep{CHARMM} in order to draw each covalent bond towards its low-energy equilibrium configuration. We denote the \textsc{FlowBack+LJ/Bonds} drift -- the \textsc{FlowBack} drift \(v_{\gamma}\) plus these three auxiliary corrections -- by \(v_{\theta}\). This is the baseline flow that will be modified by energy-based adjustment in \textsc{FlowBack-Adjoint}.  Additional algorithmic details, including velocity clamping, scaling, and force constants, are provided in \blauw{Appendix~A.1}.

These three corrections -- chirality enforcement, steric clash removal, and bond relaxation -- address the residual steric overlaps, stretched bonds, and occasional D-side chains that \textsc{FlowBack} is prone to produce, particularly for large choices of \(\sigma_{p}\), and represent an inductive bias that alleviates the degree of additional steering that must be done by the molecular mechanics energy gradients propagated by adjoint matching in \textsc{FlowBack-Adjoint}.\textsc{FlowBack+LJ/Bonds} may also be considered an ablation benchmark for \textsc{FlowBack-Adjoint} that introduces physics-based corrections on top of which we may discern the additional benefit conveyed by the learned energy-based adjoint matching procedure. 
  

%
\subsection{\textsc{FlowBack-Adjoint}}\label{subsec:fbadj}
The \textsc{FlowBack+LJ/Bonds} augmentation produces an improved (i.e., lower) energy distribution among the AA backmapped configurations relative to \textsc{FlowBack}, but still fails to reproduce that produced in a MD simulation, with an elevated mean of approximately 1-3 kcal/mol.residue (cf.\ Section~\ref{sec:results}). This discrepancy is apparently small, but thermodynamically significant, since natural proteins tend to be marginally stable with a free energy energy favoring the folded state of a typical protein molecule lying in the range 5-20 kcal/mol \citep{dill1990dominant}. Even a modest residual energy bias translates into an exponential distortion of thermodynamic probabilities through the Boltzmann factor, undermining applications such as binding-affinity ranking that depend on precise energy ordering \citep{aldeghi2019}. Consequently, eliminating steric clashes and restoring covalent geometry is necessary but not sufficient for accurate modeling of the AA backmapped ensemble. This requirement motivates the development of \textsc{FlowBack-Adjoint} to augment the velocity field with adjoint terms that nudge trajectories toward lower-energy regions of configurational space. 

Unlike \textsc{FlowBack}, which samples its initial coordinates from a Gaussian distribution of heavy atoms centered on each parent C\(_\alpha\), \textsc{FlowBack-Adjoint} generates initial heavy atom positions from a zero-centered normal prior \(x_{0} \sim q_0(x) = \mathcal{N}\!\bigl(0,\sigma_{p}^{2}I\bigr)\) and learns an additive offset relative to the parent C\(_\alpha\). The C\(_\alpha\) coordinates are still passed as an input, and the predicted atom positions are placed relative to those coordinates before the EGNN featurization as well as for all subsequent geometric and energetic evaluations. The distribution of generated structures $q_1$ is identical to one obtained with the C\(_\alpha\)-centered prior since the choice of base distribution changes only an internal coordinate representation and has no impact on any observable metrics, ensuring that all comparisons to the previous \textsc{FlowBack} model remain fully valid. The re-parameterization is required because the adjoint matching framework of \citet{domingoenrich2025adjointmatching} employed in this work assumes a zero-mean  Gaussian base distribution in order to derive an unbiased estimator between the original flow and a reward-tilted target. Adopting \(\mathcal{N}(0,\sigma_{p}^{2}I)\) priors preserves the theoretical consistency of the adjoint correction while still allowing the model to condition explicitly on the supplied backbone. For completeness we also re-train the baseline \textsc{FlowBack} and \textsc{FlowBack+LJ/Bonds} model with the same \(\mathcal{N}(0,\sigma_{p}^{2}I)\) prior. 

To train \textsc{FlowBack-Adjoint}, we provide the C$_\alpha$ positions $C$, protein topologies (i.e., covalent connectivity of the constituent atoms) $T$, atomic positions $x$, and interpolation time $t$, to an EGNN that learns the trainable weights and biases $\phi$ of the learned conditional flow $v_{\phi}(x_t, t, C, T)$. The learned flow $v_{\phi}$, represents a fine-tuned adjustment of the \textsc{FlowBack+LJ/Bonds} flow $v_\theta$ via adjoint matching. Adjoint matching is a post-training procedure that refines an existing diffusion or flow-matching generator so its outputs better satisfy a user-defined score or reward \citep{domingoenrich2025adjointmatching} by deriving a gradient correction that nudges the model's velocity field to raise the final reward under a specified objective function.
In the present work, the adjoint matching procedure modifies a flow that samples AA configurations $x$ from a base target distribution $p_{\theta}(x)$ by steering the flow to instead sample from a new reweighted distribution $p_{\phi}(x) = p_{\theta}(x)e^{R(x)}$, where $R(x)$ is the reward function. 

Tilting a generator from its base distribution \(p_{\theta}(x)\) to the new, reward-weighted density can be cast as a stochastic optimal control (SOC) problem wherein one seeks a time-dependent control \(u_t(x)\) that minimizes the expected path cost \(\mathbb{E}\!\left[\int_0^1\tfrac{1}{2}\|u_t\|^2\,\mathrm{d}t - R(x_1)\right]\) such that the $t=1$ distribution follows the reward-weighted density $p_{\phi}(x) = p_{\theta}(x)e^{R(x)}$ \citep{domingoenrich2024socmatching}. To solve this SOC problem, we follow the approach presented in \citet{domingoenrich2025adjointmatching}. A key requirement for this approach is that  the underlying diffusion must be memoryless, meaning that the variance of its driving noise satisfies \(\sigma_t^2(t) = 2\eta_t\) with \(\eta_t=(1-t)/t\). Because \(\eta_t\to\infty\) as \(t\to 0\) and \(\eta_t\to 0\) as \(t\to 1\), the process randomizes the initial draw and gradually ``de-randomizes" toward the end of the flow as the target distribution increases in importance.In practice, the divergence at \(t\to 0\) is avoided by adding a small offset in both the numerator and denominator, giving $\eta_t = (1-t+\Delta t)/(t + \Delta t)$. The memoryless schedule guarantees that the controlled dynamics reproduce \(p_{\phi}\) without bias. Violations of the memoryless requirement can leave a residual imprint of the starting noise and corrupt the tilt \citep{domingoenrich2025adjointmatching}. Adjoint Matching provides a practical solution to this SOC by computing, via an adjoint-state recursion, the minimal-variance estimator of the optimal control and realizing it as a small additive correction to the velocity field.  Applied to the memoryless diffusion above, the method drives the generator toward the reward-weighted density $p_{\phi}(x) = p_{\theta}(x)e^{R(x)}$.

To generate a heavy-atom backbone from a fixed C\(_\alpha\) trace within conditional flow-matching  framework, we model the unknown coordinates as a time-indexed random path \(\bigl(x_t\bigr)_{t\in(0,1]}\) that evolves from an easy-to-sample prior at \(t\!=\!0\) to the target distribution at \(t\!=\!1\). In conventional flow-matching, samples are advanced by the deterministic ODE
\(\mathrm{d}x_t = v_{\theta}\!\left(x_t,t\right)\mathrm{d}t\), whose associated
continuity equation \(\partial_t p_{\theta}(x_t)+\nabla\!\cdot\!\bigl(p_{\theta}(x_t)\,v_{\theta}(x_t)\bigr)=0\) 
guarantees probability conservation \citep{ma2024}. Any dynamics that satisfy this continuity
equation are admissible, and to satisfy the SOC formulation, we adopt the memoryless stochastic
variant,
\begin{equation}
\mathrm{d}x_t
= 2\,v_{\phi}\!\left(C,\,T,\,x_t,\,t\right)\mathrm{d}t
\;-\;\frac{1}{t}\,x_t\,\mathrm{d}t
\;+\;\sigma_t(t)\,\sigma_p\,\mathrm{d}W_t,
\label{eq:memoryless-flow}
\end{equation}
where \(W_t\) is a standard Brownian motion and
\(\sigma_t^2(t) = 2\eta_t\) with \(\eta_t = (1-t + \Delta t)/(t + \Delta t)\) \citep{domingoenrich2025adjointmatching}.
We then compute the adjoint costate vector \(a_t\in\mathbb{R}^{3N}\) that measures how an infinitesimal perturbation of the system state \(x_t\) would change the expected terminal reward \(\mathbb{E}[R(x_1)]\) as measured by the partial derivative \(a_t=\partial\mathbb{E}[R(x_1)]/\partial x_t\).  The adjoint therefore carries backward in time the sensitivity information needed to turn local updates of the base velocity field \(v_{\theta}\) into an update velocity field $v_\phi$ that produces a global improvement of the reward objective.  Because the underlying diffusion uses the memoryless noise schedule \(\sigma_t^2(t)=2(1-t)/t\), it has been shown that the adjoint obeys a simpler ordinary differential equation known as the ``lean adjoint ODE" \citep{domingoenrich2025adjointmatching},
\begin{equation}\label{eq:adj_ode}
\frac{\mathrm{d}a_t}{\mathrm{d}t}
   \;=\;
   -\,a_t^{\!\top}\,
     \nabla_x\!\Bigl(2\,v_{\theta}(x_t,t)-\tfrac{1}{t}x_t\Bigr),
\qquad
a_{\,1}=-\nabla_x R(x_{\,1}),
\end{equation}
which contains only the deterministic drift and a terminal condition given by the negative reward gradient. Operationally within \textsc{FlowBack-Adjoint}, we evaluate the required Jacobian-vector products using automatic differentiation and integrate this equation backwards over the same 100 explicit Euler steps used in the forward simulation. 

Having numerically solved the lean adjoint ODE for $a_t$, we now steer the flow $v_{\theta}$ toward the reward-weighted density $p_{\phi}(x) = p_{\theta}(x)e^{R(x)}$ under a time-dependent control force \(u_t(x)\), which, under the dynamics of Eqn.~\ref{eq:memoryless-flow}, becomes \(u_t(x) = - \left( \sigma_t(t)^{2} / 2 \right) a_t\) \citep{Kappen_2005}. The reward-adjusted velocity therefore becomes,
\[
v_{\phi}(x_t,t) \;=\; v_{\theta}(x_t,t)
        \;-\;\frac{\sigma_t(t)^{2}}{2}\,a_t .
\]
Re-arranging gives a residual that should vanish under perfect control,
\[
\frac{2}{\sigma_t(t)}\bigl(v_{\phi}-v_{\theta}\bigr)
        \;+\;\sigma_t(t)\,a_t \;=\; 0,
\]
which informs the adjoint matching loss--scaled by a factor $\sigma_p^2$ for consistency with the flow-matching integration,
\begin{equation}\label{eq:adjointloss}
\mathcal{L}_{\text{adjoint}}
   \;=\;
   \sigma_p^2
   \sum_t
   \left\lVert
      \frac{2}{\sigma_t(t)}
      \bigl(v_{\phi}(x_t,t)-v_{\theta}(x_t,t)\bigr)
      \;+\;
      \sigma_t(t)\,a_t
   \right\rVert^{2},
\end{equation}
defined as the sum of squared residuals over all time steps of the flow. Operationally, we minimize \(\mathcal{L}_{\text{adjoint}}\) over the trainable parameters $\phi$ of the EGNN encoding the flow $v_\phi$ using the Adam optimizer \citep{kingma2014adam}.

Until now, we have left the form of the reward function unspecified. In principle, any differentiable function of $x$ in Eqn.~\ref{eq:adj_ode} is permissible. In this work, we choose $R(x) = -\lambda U(x)$, where $\lambda$ is a hyperparameter controlling the degree to which the reward function modulates the base distribution and $U(x)$ is the energy assigned to an AA configuration $x$ by the CHARMM27 molecular mechanics force field \citep{CHARMM}. This choice of reward function can be conceived as promoting a reweighted distribution favoring low-energy configurations by penalizing those with high values of $U(x)$. In empirical tests, we have found that setting $\lambda$ = 0.01 provides a good balance between training stability and efficient modulation of the backmapped distribution towards low-energy AA configurations.
%

\subsection{Training}
\subsubsection{Dataset}

Training data for \textsc{FlowBack-Adjoint} was collected from the D.E.~Shaw Research (DESRES) all-atom molecular dynamics trajectories   
of 11 fast-folding mini-proteins in water ranging in size from 10-100 residues with simulation trajectories ranging in length from 104 $\mu$s -- 2.9 ms \citep{DESRESTrajs}. (We discarded villin from the original 12 mini-protein ensemble due to the presence of an unnatural amino acid.) We performed uniform subsampling of each trajectory to curate approximately 2000 structures per protein and compile a training set of $\sim$22,000 AA configurations. The C$_\alpha$ coordinates of these proteins are treated as $C$, their topological information are labeled $T$, and their coordinates are denoted $x$, all of which are input into training the \textsc{FlowBack-Adjoint} models as described in Section~\ref{sec:meth}. 
We randomly select the trajectories of 9/11 mini-proteins for training -- BBA, BBL, $\lambda$-repressor fragment, NTL9, Chignolin, Trp-Cage, Protein G, UVF, and $\alpha$3D -- and hold out the remaining two -- WW Domain and Protein B -- for validation. \textsc{FlowBack} was trained on protein structures curated from the Protein Data Bank \citep{king2021sidechainnet,alquraishi2019proteinnet,10.1093/nar/28.1.235,jones2025flowback}, \textsc{FlowBack+LJ/Bonds} required no additional training beyond the addition of the three physics-based auxiliary fields, and we train \textsc{FlowBack-Adjoint} on the DESRES MD simulation data. We made this choice since the MD data are posited to better represent equilibrium Boltzmann-distributed structural ensembles, which are more appropriate for learning energy-based corrections relative to static crystallographic structures that are subject to significant structural perturbations within the protein crystals required for experimental structure determination. 

%
\subsubsection{Energy Computation}

To train FlowBack-Adjoint using the adjoint matching objective, we require the backmapped AA configurations $x$ along with their corresponding potential energies $U(x)$ and gradients (i.e., forces) $\nabla_x U(x)$. In the previous section, we described how the training samples $x$ are drawn from DESRES molecular dynamics trajectories. Here, we describe how we evaluate $U(x)$ for these structures using a CHARMM27 molecular mechanics force field \citep{CHARMM}. Since the DESRES datasets provide atomistic coordinates but not simulation parameters or energies, we first convert each configuration to a format suitable for energy evaluation. Specifically, we export each structure as a PDB file, then use \texttt{pdb2gmx} from GROMACS \citep{ABRAHAM201519} to add missing hydrogens, assign termini, and generate CHARMM27 \citep{CHARMM} topology and coordinate files. The periodic box is expanded by 2 nm in each direction to prevent boundary interactions during energy computation. We then load these topologies into OpenMM \citep{Eastman2017OpenMM7} and evaluate the bonded and non-bonded energy terms using a 1.0 nm real-space cutoff and Particle Mesh Ewald (PME) electrostatics \citep{PME}. We restrict the energy computation to include only the atoms produced by \textsc{FlowBack-Adjoint} (i.e., we exclude hydrogens and termini, while adding their partial charges onto their covalently-bonded neighboring atoms) ensuring consistency between the generated samples and the energy model. The protein configurations populating the DESRES training trajectories were generated in solvent and so sample the configurational ensemble expected in an aqueous environment, but the trajectories lack the explicit solvent coordinates. As such, energy evaluations were conducted in the gas phase. The values of the energy $U(x)$ and forces $\nabla_x U(x)$ were then used as reward signals in the adjoint matching objective (Section~\ref{subsec:fbadj}). 



%
\subsubsection{Training Loop}
Adjoint matching provides a lightweight and effective fine-tuning scheme for generative flows. To train \textsc{FlowBack-Adjoint}, we simulate memoryless flow trajectories (Eqn.~\ref{eq:memoryless-flow}), use these trajectories to compute adjoint states by solving the lean adjoint ODE (Eqn.~\ref{eq:adj_ode}) using Euler's method, and finally adjust the parameters of $v_\phi$ to minimize the adjoint loss function (Eqn.~\ref{eq:adjointloss}). Unlike the original \textsc{FlowBack} model, in which the value of the noise $\sigma_p$ employed within the prior Gaussian distribution $q_0(x)$ need not be identical in training and inference, \textsc{FlowBack-Adjoint} requires that this parameter be fixed during training and remain unchanged during inference because the learned adjoint-based fine-tuning $v_\theta \to v_\phi$ is calibrated to trajectories drawn from the fixed prior $\mathcal{N}(0,\sigma_p^{2}I)$.
By backpropagating energy gradients through the sampled trajectories via the lean adjoint ODE, the reward-tilted flow adjusts its dynamics to achieve better downstream performance without retraining the model from scratch. The complete training loop just described herein is presented in pseudo-code in Algorithm~\ref{alg:amtraining}. An accounting of all of the \textsc{FlowBack-Adjoint} parameters and hyperparameters is detailed in \blauw{Appendix A.1} and training curves are presented in \blauw{Figure A.1}. 
%
\begin{algorithm}
\caption{\textsc{FlowBack-Adjoint} Training Loop}
\begin{algorithmic}\label{alg:amtraining}
\INPUT Base model $v_\theta$, fine-tuned model $v_\phi$, C$_\alpha$ coordinates $\{C\}$, protein topologies $\{T\}$, prior noise $\sigma_p$, number of Euler steps $N$
\FOR{each training step}
    \STATE Sample random training configuration $i \sim \mathcal{U}\{0, |C|\}$
    \STATE Sample initial coordinates of heavy atoms from prior distribution $x_0 \sim \mathcal{N}(0, \sigma_p^2 I)$
    \STATE Compute the memoryless noise schedule regularized to avoid the divergence at $t \to 0$
    $$\sigma_t^2(t) = \frac{2(1-t+\Delta t)}{t + \Delta t}$$
    \STATE Simulate the memoryless trajectory $\{x_t\}_{t=0}^{1}$ in time-steps of $\Delta t = 1/N$ from $t=0$ to $t=1$ using Euler's method
    \begin{align*}
    x_{t+\Delta t}
      &= \bigl[2 v_{\phi}(C_i,T_i,x_t,t)-\tfrac1t x_t\bigr]\Delta t
         + \sigma_t(t)\sigma_p(t)\sqrt{\Delta t}\,\epsilon_t, &
    \epsilon_t \sim \mathcal{N}(0,I).
    \end{align*}
    \STATE Compute negative energy gradient (i.e.,  atomic forces) from molecular mechanics force field $a_1 = -\nabla_x U(x_1)$
    \STATE Solve backward in time using Euler's method
    to obtain $\{a_t\}_{t=0}^{1-\Delta t}$
    $$a_{t - \Delta t} = a_t + (\Delta t) a_t^T \nabla_x (2v_\theta(x_t, t) - \frac{1}{t}x_t) $$
    \STATE Select a subset of discrete time points $\mathcal{T} \subseteq \{0, \Delta t, 2\Delta t, \dots, 1\}$ 
    \STATE Compute loss  $$\mathcal{L_\text{adjoint} } = \sigma_p^2\sum_{t \in \mathcal{T}} \left| \frac{2}{\sigma_t(t)} \left( v_{\phi}(C_i, T_i, x_t, t) - v_{\theta}(C_i, T_i, x_t, t) \right) + \sigma_t(t)\, a_t \right|^2$$
    \STATE Take gradient step with respect to parameters of $v_{\phi}$
\ENDFOR
\end{algorithmic}
\end{algorithm}

\subsection{Inference}

Inference begins by initializing a heavy-atom configuration from the prior distribution \(x_0\sim\mathcal{N}(0,\sigma_p^{2}I)\). We recall that \textsc{FlowBack-Adjoint} learns an offset of each atom from its corresponding parent C$_\alpha$ position, permitting the use of a zero-mean prior (Section~\ref{subsec:fbadj}). An Euler integration over \(N_{\text{steps}} = 100\) equally spaced time steps \(t\in[0,1]\) is then conducted using the learned drift \(v_\phi(x_t, t, C, T)\), where $C$ represents the coordinates of the C$_\alpha$ trace and $T$ is the covalent topology of the protein. We also recall that the \textsc{FlowBack-Adjoint} flow $v_\phi$ is an update of the \textsc{FlowBack+LJ/Bonds} flow $v_\theta$ and so inherently contains the chirality enforcement, steric clash removal, and bond relaxation auxiliary corrections in addition to the energy-based adjoint matching adjustments (Section~\ref{subsec:ljbonds}). The coordinates $x$ are successively advanced by Euler steps \(x\leftarrow x+v_\phi \Delta t\). The final frame $x_1$ resulting from the Euler integration is anchored via a fixed offset relative to the conditioned parent C\(_\alpha\) coordinates. By virtue of the physics-based auxiliary corrections within $v_\theta$ due to \textsc{FlowBack+LJ/Bonds} and the energy-based corrections incorporated by adjoint matching due to \textsc{FlowBack-Adjoint}, the terminal backmapped AA configurations are anticipated to be stereochemically correct (i.e., L-form), possess equilibrium bond lengths, contain very few clashes, and be significantly lower in energy relative to the \textsc{FlowBack} output. Pseudo-code detailing the inference procedure is provided in Algorithm~\ref{alg:inference}.

\begin{algorithm}
\caption{\textsc{FlowBack-Adjoint} Inference Loop}
\begin{algorithmic}\label{alg:inference}
\INPUT Fine-tuned model $v_\phi$, C$_\alpha$ coordinates $\{C\}$, protein topology $\{T\}$, prior noise $\sigma_p$, number of Euler steps $N$
\STATE Sample initial coordinates of heavy atoms from prior distribution $x_0 \sim \mathcal{N}(0, \sigma_p^2 I)$
\STATE Set $\Delta t = 1/N$
\STATE Set $x_t = x_0$
\WHILE{$t < 1.0$} 
    \STATE $x_t \gets x_t + v_\phi(x_t,t,C,T) \Delta t$
    \STATE $t \gets t + \Delta t$
\ENDWHILE
\STATE \textbf{Return} trajectory $\{x_0, x_{\Delta t}, x_{2\Delta t} \ldots x_{1}\}$
\end{algorithmic}
\end{algorithm}
\subsection{Evaluation Metrics and Model Comparisons}\label{subsec:method-eval}

We evaluate the trained \textsc{FlowBack-Adjoint} model in applications to two test sets. The first test set comprises the two held-out trajectories from the D.E. Shaw Research (DESRES) fast-folding mini-protein trajectory ensemble used to train \textsc{FlowBack-Adjoint} \citep{DESRESTrajs}: (i) the WW domain, a 35-residue triple-stranded $\beta$-sheet with a 1135 $\mu$s trajectory downsampled to 2230 frames, and (ii) Protein B, a 47-residue three-helix bundle with a 104 $\mu$s trajectory downsampled to 2602 frames. The second test set comprises a structural ensemble designed to lie significantly further outside the training distribution. We selected 19 proteins from the \textsc{BioEmu} \texttt{OOD60} benchmark that was curated specifically for low-sequence identity and uncommon folds \citep{Lewis2024}. To probe length-scale limits we added to these 19 test proteins the 97 kDa (418 residue) \textit{vibrio Cholerae} LapD c-di-GMP receptor module (PDB \texttt{6PWK}) to include a very large protein, and three single-point mutants of the DNA-binding protein DUX4 (PDB \texttt{5Z2S}) to assess sensitivity to subtle sequence changes. Together, these 23 proteins present a challenging out-of-distribution suite spanning unusual folds, long chains, and near-native variants. No sequence in this second test set shares more than 60\% sequence identity with any of the nine DESRES proteins \citep{DESRESTrajs} employed in adjoint matching training or any sequence in the original SidechainNet \citep{king2021sidechainnet} splits used to train the original \textsc{FlowBack} model. For each of the 23 proteins in this test set we used \textsc{BioEmu} to generate 1,000 backbone-only conformations and then constructed the side-chains using \textsc{FlowBack}, \textsc{FlowBack+LJ/Bonds}, \textsc{FlowBack-Adjoint}, and \textsc{HPacker} as \textsc{BioEmu}'s default choice for side chain packing. 

We assess model performance over these two test sets using a variety of tests designed to probe different aspects of the structure and energy of the backmapped AA configurational ensemble.
Tests 1-3 -- Bond Score, Clash Score, and Diversity Score -- follow the metrics introduced in \citet{jones2023diamondback}, Test 4 -- Energy Divergence -- quantifies the overlap in the CHARMM27 energy distribution of the backmapped configurational ensemble with the ground-truth MD trajectories, and Test 5 -- MD Test -- assesses the stability of MD simulations launched from the backmapped AA configurations without any additional energy relaxation. All five tests are conducted for \textsc{FlowBack}, \textsc{FlowBack+LJ/Bonds}, and \textsc{FlowBack-Adjoint}.

 $\bullet$ \textbf{Test 1 -- Bond Score ($\uparrow$).} We gauge physical plausibility of the covalent bond network by computing the fraction of bonds whose lengths are within 10\% of the amino-acid reference. A 100\% bond quality score is ideal.

$\bullet$ \textbf{Test 2 -- Clash Score ($\downarrow$).} We quantify steric clashes as the share of residues that sit within 1.2 \r{A} of any other residue. A clash score of 0\% is ideal.

$\bullet$ \textbf{Test 3 -- Diversity Score ($\downarrow$).} This metric evaluates how well the all-atom (AA) reference structure is represented within the ensemble of AA structures generated from a single C$_\alpha$ trace, and is defined as $DIV = 1 - RMSD_\mathrm{gen}/RMSD_\mathrm{ref}$, where $RMSD_\mathrm{gen}$ is the mean pairwise heavy atom root mean squared deviation (RMSD) between an ensemble of $G$ backmapped AA configurations generated from the same C$_\alpha$ trace, and $RMSD_\mathrm{ref}$ is the mean heavy atom RMSD of these $G$ configurations from the ground-truth reference configuration \citep{jones2023diamondback}. Since CG to AA backmapping is inherently one-to-many, a desirable backmapping algorithm should yield a physically plausible, diverse ensemble that includes the reference conformation among its possibilities. A score of \(DIV = 1\) indicates a fully deterministic method (i.e., all generated structures are identical to one another), whereas values of \(DIV = 0\) indicates both high diversity of the generated ensemble and good consistency with the reference (i.e., the reference structure lies within the distribution of generated configurations). In general, the generated configurations are anticipated to exhibit a tighter distribution around their own mean than around the reference configuration (i.e., $RMSD_\mathrm{gen} < RMSD_\mathrm{ref}$ such that DIV generally lies on the $[0,1]$ interval. However, this inequality is not strict and we do observe minor violations such that $DIV < 0$, indicative of tighter dispersion around the reference than around the generated ensemble mean. For evaluation purposes, we treat small negative values of $-0.1 < DIV < 0$ as operationally equivalent to an ideal score of $DIV = 0$.


$\bullet$ \textbf{Test 4 -- Energy Divergence ($\downarrow$).} This test measures how closely the potential energy distribution of the generated ensemble matches that of the reference ensemble.  Let \(p_k^{\text{gen}}\) and \(p_k^{\text{ref}}\) denote the discretized probability distributions of observing a configuration with a CHARM27 energy in bin \(k\) under, respectively, the generated backmapped AA ensemble and the reference data harvested from the DESRES MD trajectory. We quantify the distance of the reference distribution from the generated distribution using the Kullback-Leibler (KL) divergence, 
\[
KL\!\left(p_{\text{ref}}\;\|\;p_{\text{gen}}\right)
  = \sum_{k=1}^{B} p_k^{\text{ref}}
    \ln\!\frac{p_k^{\text{ref}}}{p_k^{\text{gen}}}, 
\]
where \(B\) is the number of bins. The KL divergence, also known as the relative entropy, is non-negative statistical distance that attains its ideal minimum at \(KL = 0\) indicating that the two distributions are identical.

$\bullet$ \textbf{Test 5 -- MD Test ($\uparrow$).} This test empirically assesses the fraction of backmapped AA configurations that can be used to initialize a stable MD run in \textsc{OpenMM} \citep{Eastman2017OpenMM7} without additional energy minimization.  Commencing from \(N_{\text{start}}\) backmapped AA configurations, we count \(N_{\text{fin}}\), the number of configurations that produce a stable 20 ps gas phase MD trajectory. A success fraction $\Phi = N_{\text{fin}}/N_{\text{start}}$ of 100\% is ideal. Details of the simulation protocol are provided in \blauw{Appendix A.2}.

\section{Results}\label{sec:results}

\subsection{Hold-out DESRES Trajectories: WW domain and Protein B}

We evaluated the performance of \textsc{FlowBack-Adjoint} in an application to the two hold-out trajectories of WW domain and Protein B from the  DESRES fast-folding mini-protein trajectory ensemble \citep{DESRESTrajs}. This assessment represents a moderate out-of-sample test of \textsc{FlowBack-Adjoint} in an application to novel mini-proteins not seen during training. In Table~\ref{tab:comparison}, we present a comparison of \textsc{FlowBack}, \textsc{FlowBack+LJ/Bonds}, and \textsc{FlowBack-Adjoint} in the bond score (Test 1), clash score (Test 2), diversity score (Test 3), and energy distribution (Test 4) at for three different choices of the prior noise $\sigma_p$ (Section~\ref{subsec:method-eval}).
We recall that $\sigma_p$ controls the magnitude of the initial noise distribution in the flow-matching path and largely adjusts the trade-off between structural accuracy and diversity of the AA ensemble: models with larger $\sigma_p$ are expected to typically produce more diverse conformational ensembles but with degraded bond, clash, and energy scores.


\textbf{Structure.} In terms of structural accuracy (Tests 1 and 2) we find that for both WW domain and Protein B, \textsc{FlowBack-Adjoint} offers superior performance across in bond score and clash score. At the standard value of $\sigma_p$ = 0.003 nm recommended by the original \textsc{FlowBack} paper, \textsc{FlowBack-Adjoint} achieves 99.92\% and 99.97\% bond scores on the WW domain and Protein B, respectively, reducing the number of bond errors by 92\% and 96\% respectively compared to the base \textsc{FlowBack} model, and tying with \textsc{FlowBack+LJ/Bonds}. \textsc{FlowBack-Adjoint} also completely eliminates all steric clashes in both proteins, achieving perfect clash scores of 0\%, representing a marked improvement from the 0.2383\% and 0.2003\% levels of \textsc{FlowBack}. \textsc{FlowBack+LJ/Bonds} achieves clash scores of 0.0077\% and 0.0033\%, indicating that the Lennard-Jones corrections eliminates the preponderance of clashes, but only by incorporating the energy-based fine tuning within \textsc{FlowBack-Adjoint} can we drive this to zero.

\textbf{Diversity.} One failing of the $\sigma_p$ = 0.003 nm model is in the diversity score (Test 3), where the incorporation of the adjoint matching degrades the diversity score \textsc{FlowBack-Adjoint} relative to \textsc{FlowBack} by 13\% and 20\%, respectively, for WW domain and Protein B. This can be understood as the promotion of lower-energy structures reducing the structural diversity by attenuating the high-energy tail of the configurational ensemble produced by \textsc{FlowBack}. \textsc{FlowBack}'s solution to this diversity problem was to increase $\sigma_p$ to allow for the model to access a larger subset of configurational space \citep{jones2025flowback}. As illustrated in the results for the $\sigma_p$ = 0.005 nm and 0.010 nm \textsc{FlowBack} models, this substantially degrades the bond and clash scores, which fall to $\sim$85\% bond score and 4-8\% clash scores over the two proteins at the highest $\sigma_p$ = 0.010 nm noise required to produce a near perfect diversity scores close to zero. Critically, \textsc{FlowBack-Adjoint} alleviates this problem by breaking the trade-off between structural accuracy and configurational diversity. At $\sigma_p$ = 0.010 nm, \textsc{FlowBack-Adjoint} achieves a near perfect diversity score of $\sim$0 while maintaining bond scores of 98.63\% and 99.37\% and clash scores of 0.1166\% and 0.0278\%, respectively, for WW domain and Protein B. A visual of the comparison of the AA ensembles produced by the three models at the three different noise levels in the context of a single residue of Protein B is provided in \blauw{Figure A.2}.

\begin{table}[h]
    \centering
    \caption{Structural, diversity, and energy performance of \textsc{FlowBack}, \textsc{FlowBack+LJ/Bonds}, and \textsc{FlowBack-Adjoint} models over the two hold-out DESRES trajectories of WW domain and Protein B \citep{DESRESTrajs}. The best-performing values in each column are highlighted in \textbf{bold}. A KL divergence of $\infty$ in the energy score implies the distribution of energies over the reconstructed ensemble possesses no overlap with that from the MD trajectory. We recall that the diversity scores approximately lie on the [0,1] interval, but small violations below the lower bound are possible, and the optimal diversity score is selected as that which lies closest to zero.}

    \begin{tabular}{lccccc}
      \hline 
      \multicolumn{6}{c}{\textbf{WW domain}} \\
      \textbf{Model} & $\sigma_p$ (nm) & \textbf{Bond} ($\uparrow$) & \textbf{Clash} ($\downarrow$) & \textbf{Diversity} ($\downarrow$) & \textbf{Energy} ($\downarrow$) \\
      \hline
      \lownoise \textsc{FlowBack}           & 0.003 & 99.00\% & 0.2383\% & 0.3532 & 0.1397 \\
      \lownoise \textsc{FlowBack+LJ/Bonds}  & 0.003 & \textbf{99.93\%} & 0.0077\% & 0.3627 & 0.0074 \\
      \lownoise \textsc{FlowBack-Adjoint}   & 0.003 & 99.92\% & \textbf{0.0000\%} & 0.4010 & \textbf{0.0018} \\ 
      \midnoise \textsc{FlowBack}           & 0.005 & 97.28\% & 1.3107\% & 0.1368 & 0.2010 \\
      \midnoise \textsc{FlowBack+LJ/Bonds}  & 0.005 & 99.84\% & 0.0231\% & 0.1880 & 0.0390 \\
      \midnoise \textsc{FlowBack-Adjoint}   & 0.005 & 99.86\% & 0.0051\% & 0.2300 & 0.0137 \\ 
      \highnoise \textsc{FlowBack}          & 0.010 & 84.79\% & 7.6790\% & -0.0835 & $\infty$ \\
      \highnoise \textsc{FlowBack+LJ/Bonds} & 0.010 & 98.31\% & 0.2755\% & -0.0624 & 0.2030 \\
      \highnoise \textsc{FlowBack-Adjoint}  & 0.010 & 98.63\% & 0.1166\% & \textbf{-0.0270} & 0.2010 \\ 
      \hline
    \end{tabular}

    \vspace{0.9em}

    \begin{tabular}{lccccc}
      \hline 
      \multicolumn{6}{c}{\textbf{Protein B}} \\
      \textbf{Model} & $\sigma_p$ (nm) & \textbf{Bond} ($\uparrow$) & \textbf{Clash} ($\downarrow$) & \textbf{Diversity} ($\downarrow$) & \textbf{Energy} ($\downarrow$) \\
      \hline
      \lownoise \textsc{FlowBack}           & 0.003 & 99.30\% & 0.2003\% & 0.3117 & 0.0859 \\ 
      \lownoise \textsc{FlowBack+LJ/Bonds}  & 0.003 & \textbf{99.97\%} & 0.0033\% & 0.3317 & 0.0087 \\
      \lownoise \textsc{FlowBack-Adjoint}   & 0.003 & \textbf{99.97\%} & \textbf{0.0000\%} & 0.3730 & \textbf{0.0017} \\
      \midnoise \textsc{FlowBack}           & 0.005 & 97.64\% & 0.6517\% & 0.1185 & 0.1851 \\
      \midnoise \textsc{FlowBack+LJ/Bonds}  & 0.005 & 99.95\% & 0.0016\% & 0.1561 & 0.0394 \\
      \midnoise \textsc{FlowBack-Adjoint}   & 0.005 & 99.95\% & 0.0000\% & 0.1915 & 0.0122 \\ 
      \highnoise \textsc{FlowBack}          & 0.010 & 85.29\% & 4.3543\% & -0.1112 & $\infty$ \\
      \highnoise \textsc{FlowBack+LJ/Bonds} & 0.010 & 99.26\% & 0.0499\% & -0.0711 & 0.1987 \\
      \highnoise \textsc{FlowBack-Adjoint}  & 0.010 & 99.37\% & 0.0278\% & \textbf{-0.0347} & 0.1942 \\ 
      \hline
    \end{tabular}

    \label{tab:comparison}
\end{table}

\begin{figure}[h]
    \centering
    \includegraphics[width=0.9\linewidth]{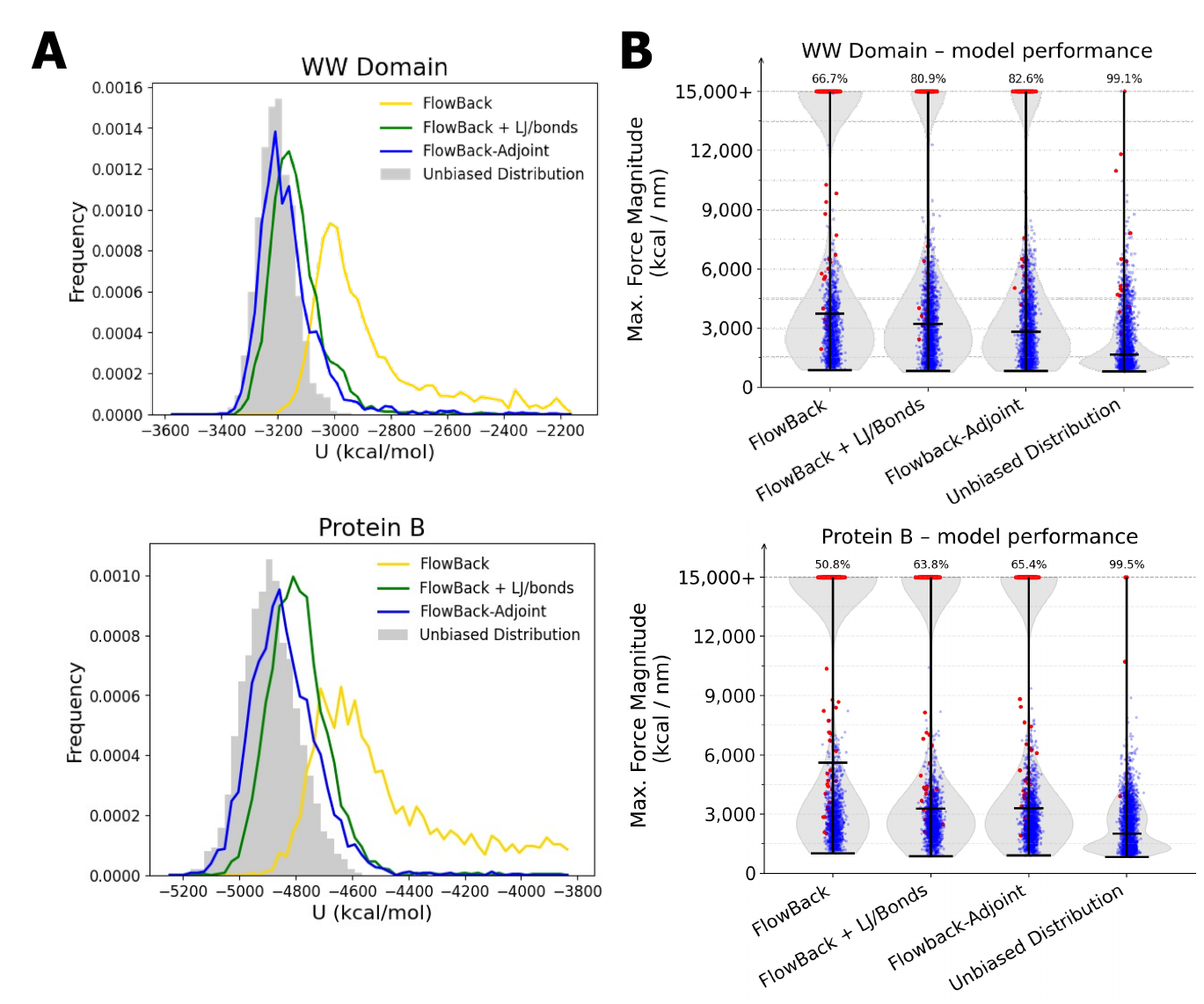}
    \caption{Energy and force performance of the \textsc{FlowBack}, \textsc{FlowBack+LJ/Bonds}, and \textsc{FlowBack-Adjoint} models over the two hold-out DESRES trajectories of WW domain and Protein B \citep{DESRESTrajs}. All models employ a noise value of $\sigma_p$ = 0.003 nm. (A) Distribution of potential energies over an ensemble of 2230 (WW domain) or 2602 (Protein B) backmapped AA configurations from each model (yellow, green, blue) relative to the distribution over the MD simulation trajectory (gray). (B) Swarm and violin plots of the distribution of the maximum force magnitudes experienced by any atom of a backmapped AA configuration before a simulation begins. Statistics for each model are aggregated over 2230 (WW domain) or 2602 (Protein B) initial AA configurations and the maximum force identified on any atom by these initial configurations when subjected to the CHARMM27 force field \citep{CHARMM}. For clarity of exposition, large forces in excess of 15,000 kcal/nm are collapsed together at the top of the plot. Test runs in which the MD simulation remained stable over the course of the 20 ps run (i.e., the simulation doesn't crash due to energy, position, or velocity values exceeding floating point limits) have their corresponding points in the swarm colored blue and those which destabilized are colored red. The lowest and median maximum atom-wise force in each swarm plot are indicated by horizontal lines and the overall success rate of stable MD simulation runs is noted as a percentage at the top of each violin.}
    \label{fig:energies}
\end{figure}

\textbf{Energy.} In terms of the energy assessment (Test 4), we find that \textsc{FlowBack-Adjoint} more faithfully captures the potential energy distribution of the hold-out MD trajectory ensembles. Focusing on the standard noise value of $\sigma_p$ = 0.003 nm, the KL divergence between the reconstructed and MD potential energy distributions drops by an order of magnitude in going from \textsc{FlowBack} to \textsc{FlowBack+LJ/Bonds} (WW domain: $0.1397 \rightarrow 0.0074$; Protein B: $0.0859 \rightarrow 0.0087$), and a further approximately 4-5-fold reduction is achieved in moving to \textsc{FlowBack-Adjoint} (WW domain: $0.0074 \rightarrow 0.0018$; Protein B: $0.0087 \rightarrow 0.0017$). Figure~\ref{fig:energies}A provides a graphical illustration of this improvement by juxtaposing the potential energy distributions generated by each model against those from the MD trajectories. Adding the Lennard-Jones and bonded corrections to the flow-matching dynamics within \textsc{FlowBack+LJ/Bonds} alleviates the majority of the energy problem, shifting the distribution toward markedly lower energies that are in much better agreement with the MD results. However, a subtle yet significant discrepancy persists until the energy-based correction is applied, with the \textsc{FlowBack-Adjoint} distribution lying very close to that from the DESRES simulation data. Nevertheless, a significant high-energy tail remains in the \textsc{FlowBack-Adjoint} distribution, indicating that matching is imperfect and that the ensemble does not fully mimic the (putative) Boltzmann distribution exhibited by the simulation data. (Although we do note that the CHARMM22$^\ast$ force field employed in the DESRES simulations \citep{DESRESTrajs} is not identical to the  CHARMM27 force field \citep{CHARMM} employed in the adjoint matching routines.) As illustrated in \blauw{Figure A.3}, similar trends persist at a noise levels of $\sigma_p$ = 0.005 nm and 0.01 nm, albeit with overall poorer matching of the reference MD energy distribution.


We recall that the adjoint matching protocol is designed to promote low-energy configurations by modulating sampling away from a base distribution $p_{\theta}(x)$ to a new distribution  $p_{\phi}(x) = p_{\theta}(x)e^{R(x)}$, where $R(x) = -\lambda U(x)$ is our reward function promoting low-energy configurations $x$ by penalizing those with high values of $U(x)$. For sufficiently strong rewards, one might anticipate mode seeking behaviors wherein the structural ensemble collapses to local energy minima causing the AA configurational ensemble to contain only low-energy ``inherent structures'' \citep{stillinger1999exponential}. This behavior would be manifest in a shifting of the \textsc{FlowBack-Adjoint} to the left (i.e., comparatively lower-energy values) relative to the MD ensemble. For the value of $\lambda$ = 0.01 employed in this work, however, we do not observe this pathology and there appears to be no catastrophic forgetting of the structural performance contained in the original \textsc{FlowBack} model. It is an interesting question for future work to explore whether the reward function might be replaced with one that seeks not to simply lower energy, but promote the Boltzmann distribution at a particular temperature of interest. 

To further probe the influence of the adjoint matching upon the energy distribution, we sought to determine whether the energies could be driven to even lower values by fitting customized single-protein \textsc{FlowBack-Adjoint} models for each of the WW domain and for Protein B and comparing their predictions to the general-purpose model trained over the nine DESRES training trajectories. Training was conducted for the same number of optimization steps used for the nine-trajectory model and the trained models then used to generate in-sample AA configurations for WW domain and for Protein B. The energy distributions for the two single-protein models show no statistically significant reduction in energies relative to the multi-protein model at an $\alpha$ = 0.05 significance threshold under a Kolmogorov-Smirnov test (\blauw{Figure A.4}; $p$ = 0.65 (WW domain) and $p$ = 0.36 (Protein B)). This observation suggests that the multi-protein \textsc{FlowBack-Adjoint} model has achieved a excellent out-of-sample performance in generating low-energy AA configurations that cannot be significantly improved by dedicated training on the specific protein of interest.

\begin{figure}
    \centering
    \includegraphics[width=\textwidth]{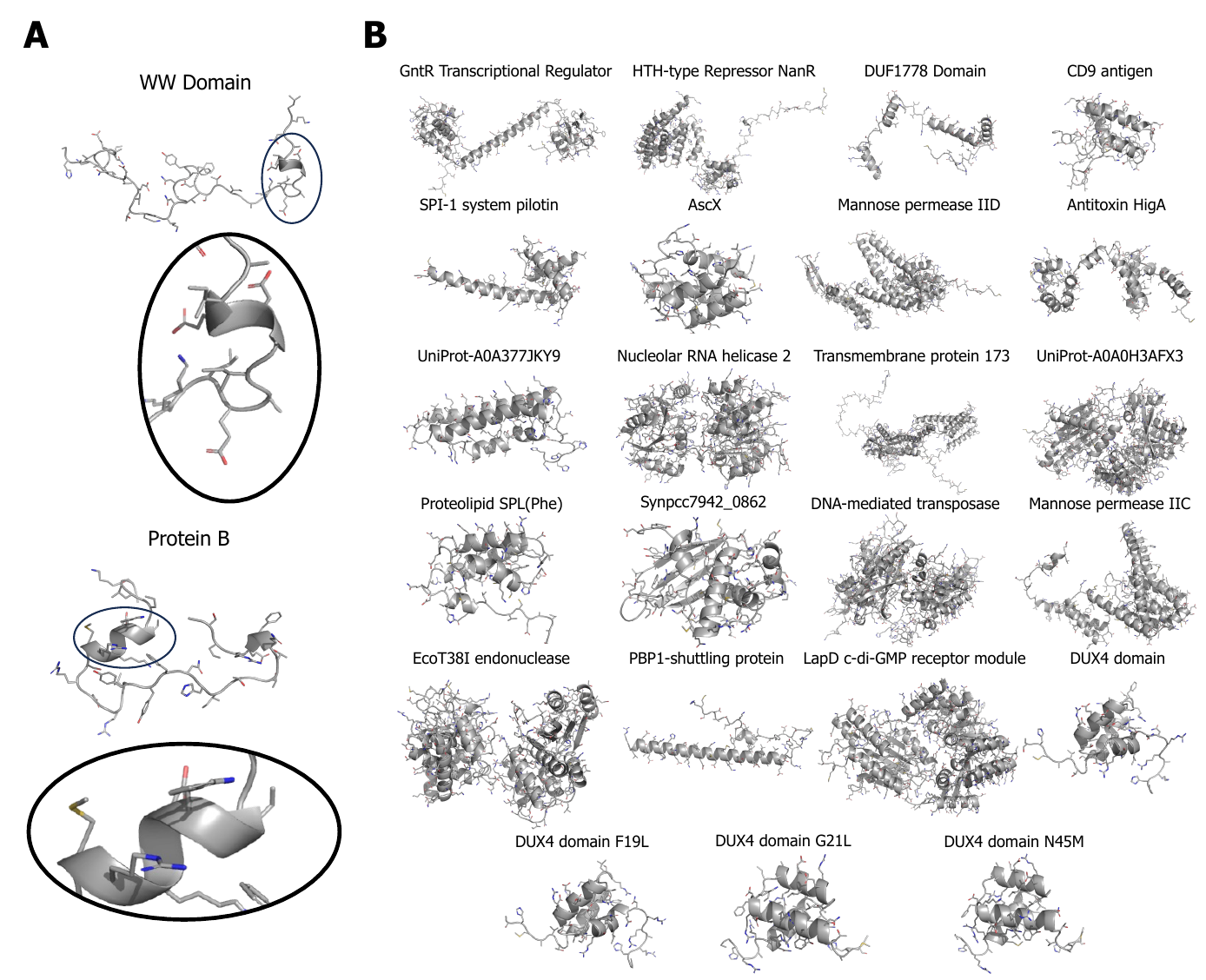}
    \caption{Representative structures of each test protein backmapped using \textsc{FlowBack-Adjoint} with $\sigma_p = 0.003$ nm. The backbone is represented as a ribbon structure and the side chains in stick representation. (A) The WW domain and Protein B from the DESRES test set with insets zooming in on their most densely packed regions. (B) AA backmapped structures of all 23 out-of-distribution proteins from \textsc{BioEmu}'s \texttt{OOD60} benchmark.}
    \label{fig:all_proteins}
\end{figure}

\textbf{Forces.} Turning, finally, to an assessment of forces (Test 5), we observe that \textsc{FlowBack-Adjoint} produces structures with far fewer large forces that can destabilize MD integrators. As illustrated in Figure~\ref{fig:energies}B, conformations generated by \textsc{FlowBack-Adjoint} exhibit smaller instantaneous force magnitudes during the validation MD run (median maximum atom-wise force = 2820 kcal/nm for WW domain and 3297 kcal/nm for Protein B) relative to \textsc{FlowBack} (3729 kcal/nm and 5588 kcal/nm) and smaller or comparable to \textsc{FlowBack+LJ/Bonds} (3217 kcal/nm and 3262 kcal/nm). Moreover, a higher proportion of MD simulation trajectories initialized with backmapped AA configurations without any additional energy minimization remain stable over a 20 ps simulation trajectory using \textsc{FlowBack-Adjoint} (82.6\% for WW domain and 65.4\% for Protein B) relative to \textsc{FlowBack} (66.7\% and 50.8\%) or \textsc{FlowBack+LJ/Bonds} (80.9\% and 63.8\%). Representative frames of backmapped structures of the WW domain and Protein B under \textsc{FlowBack-Adjoint} are visualized in Figure~\ref{fig:all_proteins}A, with close-up portrayals of the most densely packed regions of the protein, and corresponding visualizations of the Euler integration of the flow-matching trajectories are presented in \blauw{Movie A.1} and \blauw{Movie A.2}.

Nevertheless, a substantial gap remains relative to the MD simulation data, which possesses a median maximum atom-wise force = 1644 kcal/nm for WW domain and 2001 kcal/nm for Protein B, and corresponding stable simulation success rates of 99.1\% and 99.5\%. This residual discrepancy appears not to be a failing of the adjoint matching procedure, but rather a consequence of limitations in hydrogen atom placement on top of the AA backmapped structures produced by \textsc{FlowBack-Adjoint}, which, like the original \textsc{FlowBack} model, lack explicit hydrogen atoms. Analysis of our trajectories reveal that the large magnitude atom-wise forces tend to be associated with clashing or bonded hydrogen interactions, suggesting that the instabilities may be introduced by suboptimal H atom placement using the \texttt{gmx pdb2gmx} function \citep{ABRAHAM201519}, and is also the reason the MD runs initialized with MD trajectory frames do not reach a 100\% success rate. As illustrated in \blauw{Figure A.5}, if we neglect these forces from our analysis, the median maximum atom-wise forces drop substantially for all four cases -- \textsc{FlowBack}: 1557 kcal/nm for WW domain and 1946 kcal/nm for Protein B, \textsc{FlowBack+LJ/Bonds}: 1188 kcal/nm and 1264 kcal/nm, and \textsc{FlowBack-Adjoint}: 1159 kcal/nm and 1214 kcal/nm -- with both \textsc{FlowBack-Adjoint} and \textsc{FlowBack+LJ/Bonds} achieving a value commensurate with the values of 1212 kcal/nm and 1271 kcal/nm computed for the MD data. After deleting the hydrogen atoms and termini, reassigning their partial charges to the adjacent heavy atoms, and rerunning the simulation with these merged groups treated as united atoms that experience exactly the same bonded forces as their original forms, all trajectories initiated from the MD snapshots remained stable to attain a 100 \% success rate. This ideal behavior is closely approached by all three models -- 96.1\% for WW domain and 97.0\% for Protein B under \textsc{FlowBack}, 99.4\% and 99.9\% under \textsc{FlowBack+LJ/Bonds}, and 99.6\% and 99.9\% under \textsc{FlowBack-Adjoint}. Improved algorithms for hydrogen atom placement are therefore anticipated to essentially completely close the force and stability gap of \textsc{FlowBack-Adjoint}.

%
\subsection{\textsc{BioEmu} 23-protein Test Set}

We next asked whether the advantages of \textsc{FlowBack-Adjoint} carry over to proteins that lie far outside the training distribution. To test this, we inserted our integrator into the \textsc{BioEmu} pipeline \citep{Lewis2024}, which generates backbone-only equilibrium ensembles and then rebuilds side-chains using \textsc{HPacker} as a two-stage model that predicts $\chi$-dihedral angles with a lightweight, rotationally equivariant convolutional neural network and then refines atomic coordinates by minimizing losses between predicted and known $\chi$-angles \citep{visani2023}. 
Like the original \textsc{FlowBack} model, \textsc{HPacker} optimizes geometric accuracy as opposed to potential energy: it is trained to minimize root-mean-square deviation (RMSD) of side-chain heavy atoms and therefore lacks an explicit treatment of long-range electrostatics, van der Waals repulsion, or bonded strain. By replacing \textsc{HPacker} with \textsc{FlowBack-Adjoint}, we sought to determine whether the model could generate lower energy AA configurational ensembles within the \textsc{BioEmu} pipeline.

We conducted this assessment over a 23-protein benchmark comprising the 19 \textsc{BioEmu} \texttt{OOD60} targets, the c-di-GMP receptor LapD, and three point mutants of DUX4. Representative structures backmapped with \textsc{FlowBack-Adjoint} are shown in Figure~\ref{fig:all_proteins}B. We confirmed via a BLAST search that none of these proteins shared more than 60\% identity with any entry in the SidechainNet training corpus \citep{king2021sidechainnet} that was used to train the original \textsc{FlowBack} model or any of the nine DESRES proteins \citep{DESRESTrajs} used to train \textsc{FlowBack-Adjoint}, thereby guarding against any observed improvements over this test set resulting from data leakage or high sequence similarity with the training set (\blauw{Appendix A.3}). For each protein we used \textsc{BioEmu} to generate 1,000 backbone-only conformations, within which \textsc{BioEmu}'s built-in geometry filter then discarded 3-250 nonphysical traces per sequence. We then generated for each structure full AA backmapped configurations using \textsc{FlowBack}, \textsc{FlowBack+LJ/Bonds}, \textsc{FlowBack-Adjoint}, and \textsc{HPacker}. Importantly, the three FlowBack-based pipelines were given nothing beyond the C$_\alpha$ trace for each filtered conformation -- the minimal input on which these models were originally trained -- whereas \textsc{HPacker}, following its standard workflow, enjoyed the richer information content of the backbone N, C$_\alpha$, C, and O atoms. 

\begin{figure}
    \centering
    \includegraphics[width=0.8\textwidth]{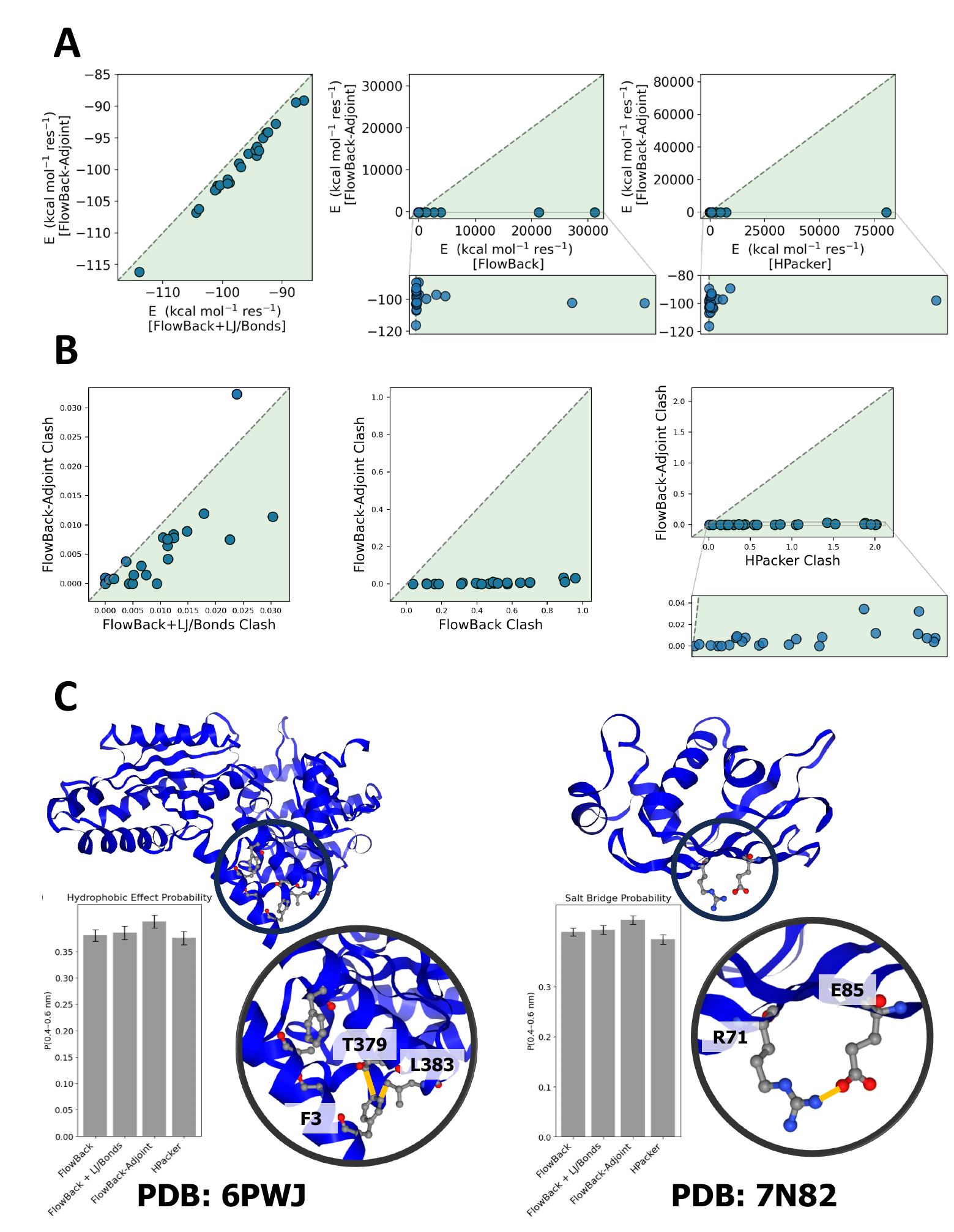}
    \caption{Performance of \textsc{FlowBack-Adjoint} as a modular replacement for side-chain reconstruction within the \textsc{BioEmu} backmapping workflow. Comparison of \textsc{FlowBack-Adjoint} (A) potential energies and (B) clash scores relative to \textsc{FlowBack}, \textsc{FlowBack+LJ/Bonds}, and \textsc{HPacker} for each of the 23 test proteins. Backbones are generated with \textsc{BioEmu} and backmapped to AA resolution by each method. All \textsc{FlowBack}-based models employ a noise level of $\sigma_p$ = 0.003 nm. Points pertain to averages over 1000 backbone-only structures generated by \textsc{BioEmu}, where for the energies we report medians and for the clash scores we report means. The green-shaded region of each plot indicates superior \textsc{FlowBack-Adjoint} performance (i.e., lower energies or fewer clashes). (C) Assessment of the prevalence of AA backmapped structures containing biologically important features in two test proteins: the heavy-atom contact occupancies between F3 of the S-helix and L383 of the EAL domain in LapD as a key hydrophobic interaction in receptor signaling \citep{Kitts2019} (left) and the C$_\zeta$-C$_\delta$ distance distribution of the R71-E85 salt bridge in the Se0862 cyanobacterial protein \citep{JOHANSSON1997859} (right). The prevalence of each motif over the 1000 AA backmapped structures for each of the four methods is displayed in the accompanying bar charts.}
    \label{fig:newproteins}
\end{figure}

Figure~\ref{fig:newproteins}A summarizes the performance of \textsc{FlowBack-Adjoint} against the three other approaches in terms of the median potential energy computed over the AA reconstructions from 1000 \textsc{BioEmu} backbone-only structures. The full energy distributions over the 1000 candidates are presented in \blauw{Figure~A.6}. For all 23 test proteins, \textsc{FlowBack-Adjoint} delivers AA ensembles with lower energies. Relative to the geometry-centric \textsc{HPacker}, the improvement is dramatic: median energies drop by roughly two orders of magnitude and eliminate the extreme high-energy structures that can plague \textsc{HPacker} ensembles. Relative to \textsc{FlowBack}, the incorporation of Lennard-Jones and bonded interaction inductive biases within \textsc{FlowBack+LJ/Bonds} cuts median energies between 6-85\%, and \textsc{FlowBack-Adjoint} realizes an additional 1-3\% reduction. Although small in a relative sense, this reduction can amount to 1-3 kcal/mol.residue, and the gains are especially pronounced for larger protein systems. Overall, \textsc{FlowBack-Adjoint} lowers single-point energies in the \texttt{OOD60} proteins by a median of $\sim$78 kcal/mol.residue relative to \textsc{FlowBack}.

Figure~\ref{fig:newproteins}B presents the mean clash scores over the 1000 structures for each of the 23 proteins in the test set to provide a complementary measure of steric quality. Consistent with the energy analysis, \textsc{FlowBack-Adjoint} outperforms both \textsc{FlowBack} and \textsc{HPacker} on every protein in the test set. On average, \textsc{FlowBack-Adjoint} reduces clashes by 98.8\% compared to the \textsc{FlowBack} prior. It also outperforms \textsc{FlowBack+LJ/Bonds} on 20/23 proteins, but there are three cases in which \textsc{FlowBack+LJ/Bonds} delivers marginally lower mean clash scores. Analysis of these three cases reveals the origin of these marginally better performance to be due to a small number of conformations with poorly-packed side-chains that skew the \textsc{FlowBack-Adjoint} mean clash scores to high values (\blauw{Figure A.7}). The observed reversals therefore point to a handful of edge cases in which the energy-based adjoint matching refinement perturbs side-chain orientations into slightly less favorable local geometries. Although these instances are rare, they highlight a potential avenue for future refinement of \textsc{FlowBack-Adjoint} by, for example, incorporating a more diverse training corpus.

As a final test, we asked whether \textsc{FlowBack-Adjoint} captures biologically meaningful features that its predecessors may overlook. 
Figure~\ref{fig:newproteins}C highlights two examples. In LapD (left), we track the heavy-atom contact occupancy between F3 of the S-helix and L383 of the EAL domain as a hydrophobic interaction known to be critical for receptor signaling \citep{Kitts2019}. In the cyanobacterial protein Se0862 (right), we examine the C$_\zeta$--C$_\delta$ distance distribution of the R71-E85 salt bridge \citep{JOHANSSON1997859}. In both the hydrophobic contact and the salt bridge, \textsc{FlowBack-Adjoint} reproduces the native interaction with higher probability over the 1000 trials relative to \textsc{FlowBack}, \textsc{FlowBack+LJ/Bonds}, and \textsc{HPacker}. It is known that both of these interactions are favorable at room temperature \citep{Pace2011-vh, Geney2006}, suggesting that they are quite energetically favorable and providing a rationale for why the energy-based inductive bias within \textsc{FlowBack-Adjoint} realizes these with higher prevalence than the other models.


Taken together, these results demonstrate that the adjoint matching refinement not only stabilizes proteins drawn from the training distribution but also generalizes to challenging, unseen sequences, delivering structurally plausible and energetically stable AA configurational ensembles across a wide range of benchmark proteins.

\section{Conclusions}

We have introduced \textsc{FlowBack-Adjoint} as a fine-tuned sophistication of the deep generative conditional flow-matching \textsc{FlowBack} model for AA backmapping. \textsc{FlowBack-Adjoint} incorporates molecular mechanics physics-based corrections to bond lengths and Lennard-Jones interactions, and an energy-based inductive bias that propagates the gradients of a molecular mechanics force field through the model using adjoint matching. The adjoint matching procedure modifies the \textsc{FlowBack} flow to promote low-energy configurations in the target ensemble without inducing catastrophic forgetting or retraining of the model from scratch. We train \textsc{FlowBack-Adjoint} over nine MD trajectories of fast-folding mini-proteins ranging in size from 10-100 residues and length from 104 $\mu$s -- 2.9 ms \citep{DESRESTrajs}. Out-of-sample tests backmapping hold-out trajectories of two additional mini-proteins and \textsc{BioEmu}-generated backbone-only traces 23 additional proteins ranging in size from 50-500 residues demonstrate \textsc{FlowBack-Adjoint} to possess state-of-the-art performance in generating structurally plausible configurations with exceedingly good bond, clash, and diversity scores, energy distributions approaching those of MD ensembles, and very few high-energy overlaps that permit nearly 100\% backmapped configurations -- subject to good hydrogen atom placement -- to be used to directly initialize stable all-atom MD simulations without energy minimization. We propose \textsc{FlowBack-Adjoint} as an accurate and efficient physics-aware deep generative model for AA backmapping from C$_\alpha$ traces, but also view the present study as a springboard for several lines of future work to ameliorate some of the model's shortcomings and further expand its scope and utility.

A first shortcoming of the current \textsc{FlowBack-Adjoint} model is that the absence of explicit solvent coordinates in our training data mean that the energy calculations used to conduct the energy-based adjoint matching are performed in the gas phase. To be clear, the protein configurations populating these trajectories were generated in the presence of solvent and so sample the configurational ensemble expected in an aqueous environment, but the energy evaluations on these configurations used to train \textsc{FlowBack-Adjoint} were conducted in the gas phase. Solvent effects play a critical role in side-chain packing, so while the current model presents an excellent demonstration of the viability of adjoint matching to tilt the backmapped AA configurational ensemble towards more stable configurations and does so using training ensembles generated in the presence of water, a model in which the energy evaluations used to perform the adjoint matching also explicitly consider solvent is an important next step. In the immediate future we plan to conduct additional fine tuning of \textsc{FlowBack-Adjoint} employing energy calculations conducted using implicit or explicit solvent models. 

A second challenge is that while \textsc{FlowBack-Adjoint} achieves very good bond lengths and low clash scores, it still produces occasional chirality inversions or $\chi$-angle outliers. About 0.04\% of reconstructed residues still possess an incorrect D-chirality, although this is significantly lower than the 1-3\% of residues that do so without the chirality correction. An ablation study reveals that nearly the entire residual energy gap between \textsc{FlowBack-Adjoint} and the unbiased MD ensemble originates from the chirality correction velocity added to the flow. When the model is trained without this correction the KL divergences between the \textsc{FlowBack-Adjoint} and MD reference energy distributions fall from 0.0018 $\rightarrow$ 0.0004 for WW domain and 0.0017 $\rightarrow$ 0.0004 for Protein B (\blauw{Figure~A.8}). This intriguing result indicates that without the chirality correction, the \textsc{FlowBack-Adjoint} model appears to be producing AA configurations drawn from a very good approximation to the Boltzmann distribution, and it is the rather \textit{ad hoc} nature of the chirality correction that is responsible for the spurious high-energy tails in the chirality-corrected distribution. This suggests that improving upon the current chirality correction procedure may produce superior energy distributions lying closer to the Boltzmann distribution. We propose transitioning from the $E(3)$-equivariant EGNNs employed in this work to chirality-aware $SE(3)$-equivariant EGNNs that may allow the learned vector field to correct stereochemical errors directly, rather than relying on \textit{post hoc} adjustments to the flow.

A third deficiency of \textsc{FlowBack-Adjoint} is its inference speed. Due to the incorporation of the Lennard-Jones and bonded corrections in the improved flow-matching integrator, the run-time of the trained model is increased by approximately three-fold relative to the original \textsc{FlowBack} model and is approximately 23-fold slower than \textsc{AttnPacker}, which stands as one of the most computationally efficient side-chain packing models. For example, backmapping the 83 CASP13 protein structures takes 1588 s for \textsc{FlowBack-Adjoint}, 507 s for \textsc{FlowBack}, and just 68 s for \textsc{AttnPacker}. We propose that a relatively straightforward means to increase \textsc{FlowBack-Adjoint}'s inference speed is to replace the physics-based bond and Lennard-Jones corrections within \textsc{FlowBack+LJ/Bonds} with an additional phase of adjoint matching. This would eliminate the need for the computation of physics-based corrective velocities that are responsible for a large fraction of the inference time. Another simple improvement may be to reduce the number of steps in the integration of the conditional flow, perhaps by using more sophisticated integrators that permit larger step sizes.

A fourth clear direction for innovation would be to develop \textsc{FlowBack-Adjoint} models capable of operating at different coarse-grained resolutions. Since the model currently operates only on C$_\alpha$ traces, it can be employed with higher-resolution CG representations by first downsampling them to C$_\alpha$ resolution, but this, of course, discards potentially very useful information for higher accuracy backmapping. One possible means to do so would be to modulate the choice of atoms participating in the CG conditioning during training following a strategy deployed in \textsc{BackDiff} \citep{liu2023backdiff} to train a transferable model capable of operating on a variety of CG representations.

A fifth avenue for exploration is the replacement of the classical CHARMM27 force field \citep{CHARMM} used throughout this work for the energy-based adjoint matching and energy distribution evaluation with a machine learned interatomic potential (MLIP) with higher accuracy -- in some cases approaching quantum mechanical accuracy -- and generalizability to arbitrary molecular systems beyond proteins and biomolecules. For example, Meta's recently released Universal Model for Atoms (UMA) potential provides energies and forces for arbitrary combinations of 83 elements through a single network trained on over $10^8$ crystal, protein, ligand, and materials structures \citep{wood2025uma}. Since \textsc{FlowBack-Adjoint} interacts with its environment only through energy gradients, replacing our CHARMM27 energy calculations with UMA and exposing the model to diverse training data beyond proteins would enable us to extend \textsc{FlowBack-Adjoint} to other classes of biomolecules and macromolecules, including nucleic acids, peptoids, lipids, peptoids, organic semiconductors, and synthetic polymers. 

A sixth opportunity lies in the generation of equilibrium molecular ensembles directly from energy functions without a pre-trained flow model \citep{havens2025}. This approach, termed adjoint sampling, reframes diffusion inference as an optimal control problem and builds upon the ideas of adjoint matching \citep{domingoenrich2025adjointmatching} but eliminates the need for paired data: any differentiable potential -- classical, neural-network, or \emph{ab initio} -- may serve as the sole training signal. A successful adjoint sampling approach could be particularly valuable in cases where systems are too large or too complex to acquire MD training data. Adjoint sampling therefore complements universal potentials like UMA, and, theoretically, can enable very efficient data-free structure generation sampled from the Boltzmann distribution. 

\section*{Conflict of Interest Disclosure}

A.L.F.\ is a co-founder and consultant of Evozyne, Inc.\ and a co-author of US Patent Applications 16/887,710 and 17/642,582, US Provisional Patent Applications 62/853,919, 62/900,420, 63/314,898, 63/479,378, 63/521,617, and 63/669,836, and International Patent Applications PCT/US2020/035206, PCT/US2020/050466, and PCT/US24/10805.

\section*{Acknowledgements}

This material is based upon the work supported by the National Science Foundation Graduate Research Fellowship Program under Grant No.\ 2140001 (A.B.). Any opinions, findings, and conclusions or recommendations expressed in this material are those of the author(s) and do not necessarily reflect the views of the National Science Foundation. This material is based on work supported by the National Science Foundation under Grant No.\ CHE-2152521. This work was completed in part with resources provided by the University of Chicago Research Computing Center. We gratefully acknowledge computing time on the University of Chicago high-performance GPU-based cyberinfrastructure supported by the National Science Foundation under Grant No.\ DMR-1828629.

\clearpage
\newpage


\bibliography{bibliography}

\begin{thebibliography}{72}
\providecommand{\natexlab}[1]{#1}
\providecommand{\url}[1]{\texttt{#1}}
\expandafter\ifx\csname urlstyle\endcsname\relax
  \providecommand{\doi}[1]{doi: #1}\else
  \providecommand{\doi}{doi: \begingroup \urlstyle{rm}\Url}\fi

\bibitem[Jumper et~al.(2021)Jumper, Evans, Pritzel, Green, Figurnov,
  Ronneberger, Tunyasuvunakool, Bates, {\v{Z}}{\'\i}dek, and
  Potapenko]{jumper2021highly}
John Jumper, Richard Evans, Alexander Pritzel, Tim Green, Michael Figurnov,
  Olaf Ronneberger, Kathryn Tunyasuvunakool, Russ Bates, Augustin
  {\v{Z}}{\'\i}dek, and Anna et~al. Potapenko.
\newblock {Highly accurate protein structure prediction with AlphaFold}.
\newblock \emph{Nature}, 596\penalty0 (7873):\penalty0 583--589, 2021.

\bibitem[Baek et~al.(2021)Baek, DiMaio, Anishchenko, Dauparas, Ovchinnikov,
  Lee, Wang, Cong, Kinch, Schaeffer, Mill{\'a}n, Park, Adams, Glassman,
  DeGiovanni, Pereira, Rodrigues, van Dijk, Ebrecht, Opperman, Sagmeister,
  Buhlheller, Pavkov-Keller, Rathinaswamy, Dalwadi, Yip, Burke, Garcia,
  Grishin, Adams, Read, and Baker]{rosettafold}
Minkyung Baek, Frank DiMaio, Ivan Anishchenko, Justas Dauparas, Sergey
  Ovchinnikov, Gyu~Rie Lee, Jue Wang, Qian Cong, Lisa~N. Kinch, R.~Dustin
  Schaeffer, Claudia Mill{\'a}n, Hahnbeom Park, Carson Adams, Caleb~R.
  Glassman, Andy DeGiovanni, Jose~H. Pereira, Andria~V. Rodrigues, Alberdina~A.
  van Dijk, Ana~C. Ebrecht, Diederik~J. Opperman, Theo Sagmeister, Christoph
  Buhlheller, Tea Pavkov-Keller, Manoj~K. Rathinaswamy, Udit Dalwadi, Calvin~K.
  Yip, John~E. Burke, K.~Christopher Garcia, Nick~V. Grishin, Paul~D. Adams,
  Randy~J. Read, and David Baker.
\newblock Accurate prediction of protein structures and interactions using a
  three-track neural network.
\newblock \emph{Science}, 373\penalty0 (6557):\penalty0 871--876, 2021.

\bibitem[Lin et~al.(2023)Lin, Akin, Rao, Hie, Zhu, Lu, Smetanin, Verkuil,
  Kabeli, Shmueli, dos Santos~Costa, Fazel-Zarandi, Sercu, Candido, and
  Rives]{esmfold}
Zeming Lin, Halil Akin, Roshan Rao, Brian Hie, Zhongkai Zhu, Wenting Lu, Nikita
  Smetanin, Robert Verkuil, Ori Kabeli, Yaniv Shmueli, Allan dos Santos~Costa,
  Maryam Fazel-Zarandi, Tom Sercu, Salvatore Candido, and Alexander Rives.
\newblock Evolutionary-scale prediction of atomic-level protein structure with
  a language model.
\newblock \emph{Science}, 379\penalty0 (6637):\penalty0 1123--1130, 2023.

\bibitem[Wu et~al.(2022)Wu, Ding, Wang, Shen, Zhang, Luo, Su, Wu, Xie, Berger,
  Ma, and Peng]{Wu2022}
Ruidong Wu, Fan Ding, Rui Wang, Rui Shen, Xiwen Zhang, Shitong Luo, Chenpeng
  Su, Zuofan Wu, Qi~Xie, Bonnie Berger, Jianzhu Ma, and Jian Peng.
\newblock High-resolution de novo structure prediction from primary sequence.
\newblock \emph{bioRxiv preprint bioRxiv:10.1101/2022.07.21.500999}, 2022.

\bibitem[Watson et~al.(2023)Watson, Juergens, Bennett, Trippe, Yim, Eisenach,
  Ahern, Borst, Ragotte, Milles, Wicky, Hanikel, Pellock, Courbet, Sheffler,
  Wang, Venkatesh, Sappington, Torres, Lauko, De~Bortoli, Mathieu, Ovchinnikov,
  Barzilay, Jaakkola, DiMaio, Baek, and Baker]{rfdiffusion}
Joseph~L. Watson, David Juergens, Nathaniel~R. Bennett, Brian~L. Trippe, Jason
  Yim, Helen~E. Eisenach, Woody Ahern, Andrew~J. Borst, Robert~J. Ragotte,
  Lukas~F. Milles, Basile I.~M. Wicky, Nikita Hanikel, Samuel~J. Pellock,
  Alexis Courbet, William Sheffler, Jue Wang, Preetham Venkatesh, Isaac
  Sappington, Susana~V{\'a}zquez Torres, Anna Lauko, Valentin De~Bortoli, Emile
  Mathieu, Sergey Ovchinnikov, Regina Barzilay, Tommi~S. Jaakkola, Frank
  DiMaio, Minkyung Baek, and David Baker.
\newblock De novo design of protein structure and function with {RFdiffusion}.
\newblock \emph{Nature}, 620\penalty0 (7976):\penalty0 1089--1100, 2023.

\bibitem[Wei et~al.(2016)Wei, Xi, Nussinov, and Ma]{wei2016}
Guanghong Wei, Wenhui Xi, Ruth Nussinov, and Buyong Ma.
\newblock {Protein Ensembles}: How does nature harness thermodynamic
  fluctuations for life? {The} diverse functional roles of conformational
  ensembles in the cell.
\newblock \emph{Chemical Reviews}, 116\penalty0 (11):\penalty0 6516--6551,
  2016.

\bibitem[Wolf et~al.(2025)Wolf, Seute, Viliuga, Wagner, St{\"u}hmer, and
  Gr{\"a}ter]{wolf2025}
Nicolas Wolf, Leif Seute, Vsevolod Viliuga, Simon Wagner, Jan St{\"u}hmer, and
  Frauke Gr{\"a}ter.
\newblock Learning conformational ensembles of proteins based on backbone
  geometry.
\newblock \emph{arXiv preprint arXiv:2503.05738}, 2025.

\bibitem[Cagiada et~al.(2025)Cagiada, Thomasen, Ovchinnikov, Deane, and
  Lindorff-Larsen]{Cagiada2025}
Matteo Cagiada, F.~Emil Thomasen, Sergey Ovchinnikov, Charlotte~M. Deane, and
  Kresten Lindorff-Larsen.
\newblock {AF2}$\chi$: Predicting protein side-chain rotamer distributions with
  {AlphaFold2}.
\newblock \emph{bioRxiv preprint bioRxiv:10.1101/2025.04.16.649219}, 2025.

\bibitem[Berka et~al.(2010)Berka, Laskowski, Hobza, and Vondr{\'a}{\v
  s}ek]{Berka2010}
Karel Berka, Roman~A. Laskowski, Pavel Hobza, and Ji{\v r}{\'\i} Vondr{\'a}{\v
  s}ek.
\newblock Energy matrix of structurally important side-chain/side-chain
  interactions in proteins.
\newblock \emph{Journal of Chemical Theory and Computation}, 6\penalty0
  (7):\penalty0 2191--2203, 2010.

\bibitem[Bachmann et~al.(2011)Bachmann, Wildemann, Praetorius, Fischer, and
  Kiefhaber]{Bachmann2011}
Annett Bachmann, Dirk Wildemann, Florian Praetorius, Gunter Fischer, and Thomas
  Kiefhaber.
\newblock Mapping backbone and side-chain interactions in the transition state
  of a coupled protein folding and binding reaction.
\newblock \emph{Proceedings of the National Academy of Sciences of the United
  States of America}, 108\penalty0 (10):\penalty0 3952--3957, 2011.

\bibitem[Abramson et~al.(2024)Abramson, Adler, Dunger, Evans, Green, Pritzel,
  Ronneberger, Willmore, Ballard, Bambrick, Bodenstein, Evans, Hung, O'Neill,
  Reiman, Tunyasuvunakool, Wu, {\v Z}emgulyt{\.e}, Arvaniti, Beattie, Bertolli,
  Bridgland, Cherepanov, Congreve, Cowen-Rivers, Cowie, Figurnov, Fuchs,
  Gladman, Jain, Khan, Low, Perlin, Potapenko, Savy, Singh, Stecula,
  Thillaisundaram, Tong, Yakneen, Zhong, Zielinski, {\v Z}{\'\i}dek, Bapst,
  Kohli, Jaderberg, Hassabis, and Jumper]{Alphafold3}
Josh Abramson, Jonas Adler, Jack Dunger, Richard Evans, Tim Green, Alexander
  Pritzel, Olaf Ronneberger, Lindsay Willmore, Andrew~J. Ballard, Joshua
  Bambrick, Sebastian~W. Bodenstein, David~A. Evans, Chia-Chun Hung, Michael
  O'Neill, David Reiman, Kathryn Tunyasuvunakool, Zachary Wu, Akvil{\.e} {\v
  Z}emgulyt{\.e}, Eirini Arvaniti, Charles Beattie, Ottavia Bertolli, Alex
  Bridgland, Alexey Cherepanov, Miles Congreve, Alexander~I. Cowen-Rivers,
  Andrew Cowie, Michael Figurnov, Fabian~B. Fuchs, Hannah Gladman, Rishub Jain,
  Yousuf~A. Khan, Caroline M.~R. Low, Kuba Perlin, Anna Potapenko, Pascal Savy,
  Sukhdeep Singh, Adrian Stecula, Ashok Thillaisundaram, Catherine Tong, Sergei
  Yakneen, Ellen~D. Zhong, Michal Zielinski, Augustin {\v Z}{\'\i}dek, Victor
  Bapst, Pushmeet Kohli, Max Jaderberg, Demis Hassabis, and John~M. Jumper.
\newblock {Accurate structure prediction of biomolecular interactions with
  AlphaFold 3}.
\newblock \emph{Nature}, 630\penalty0 (8016):\penalty0 493--500, 2024.

\bibitem[Xu et~al.(2022)Xu, Wang, Wang, and Ma]{Xu2022}
Gang Xu, Yilin Wang, Qinghua Wang, and Jianpeng Ma.
\newblock Studying protein--protein interaction through side-chain modeling
  method {OPUS-Mut}.
\newblock \emph{Briefings in Bioinformatics}, 23\penalty0 (5):\penalty0
  bbac330, 2022.

\bibitem[Zhao and Sanner(2008)]{Yong2008}
Yong Zhao and Michel~F. Sanner.
\newblock Protein--ligand docking with multiple flexible side chains.
\newblock \emph{Journal of Computer-Aided Molecular Design}, 22\penalty0
  (9):\penalty0 673--679, 2008.

\bibitem[Buch et~al.(2010)Buch, Harvey, Giorgino, Anderson, and
  De~Fabritiis]{Buch2010}
I.~Buch, M.~J. Harvey, T.~Giorgino, D.~P. Anderson, and G.~De~Fabritiis.
\newblock High-throughput all-atom molecular dynamics simulations using
  distributed computing.
\newblock \emph{Journal of Chemical Information and Modeling}, 50\penalty0
  (3):\penalty0 397--403, 2010.

\bibitem[Jing et~al.(2024)Jing, Berger, and Jaakkola]{jing2024alphafold}
Bowen Jing, Bonnie Berger, and Tommi Jaakkola.
\newblock {AlphaFold} meets flow matching for generating protein ensembles.
\newblock \emph{arXiv preprint arXiv:2402.04845}, 2024.

\bibitem[Wang et~al.(2024)Wang, Wang, Shen, Wang, Yuan, Wu, and
  Gu]{wang2024protein}
Yan Wang, Lihao Wang, Yuning Shen, Yiqun Wang, Huizhuo Yuan, Yue Wu, and
  Quanquan Gu.
\newblock Protein conformation generation via force-guided {SE(3)} diffusion
  models.
\newblock \emph{arXiv preprint arXiv:2403.14088}, 2024.

\bibitem[Lewis et~al.(2024)Lewis, Hempel, Jim{\'e}nez-Luna, Gastegger, Xie,
  Foong, Satorras, Abdin, Veeling, Zaporozhets, Chen, Yang, Foster, Schneuing,
  Nigam, Barbero, Stimper, Campbell, Yim, Lienen, Shi, Zheng, Schulz, Munir,
  Sordillo, Tomioka, Clementi, and No{\'e}]{Lewis2024}
Sarah Lewis, Tim Hempel, Jos{\'e} Jim{\'e}nez-Luna, Michael Gastegger, Yu~Xie,
  Andrew Y.~K. Foong, Victor~Garc{\'\i}a Satorras, Osama Abdin, Bastiaan~S.
  Veeling, Iryna Zaporozhets, Yaoyi Chen, Soojung Yang, Adam~E. Foster, Arne
  Schneuing, Jigyasa Nigam, Federico Barbero, Vincent Stimper, Andrew Campbell,
  Jason Yim, Marten Lienen, Yu~Shi, Shuxin Zheng, Hannes Schulz, Usman Munir,
  Roberto Sordillo, Ryota Tomioka, Cecilia Clementi, and Frank No{\'e}.
\newblock Scalable emulation of protein equilibrium ensembles with generative
  deep learning.
\newblock \emph{Science}, eadv9817:\penalty0 DOI:10.1126/science.adv9817, 2024.

\bibitem[Leman et~al.(2020)Leman, Weitzner, Lewis, Adolf-Bryfogle, Alam,
  Alford, Aprahamian, Baker, Barlow, Barth, Basanta, Bender, Blacklock, Bonet,
  Boyken, Bradley, Bystroff, Conway, Cooper, Correia, Coventry, Das, De~Jong,
  DiMaio, Dsilva, Dunbrack, Ford, Frenz, Fu, Geniesse, Goldschmidt, Gowthaman,
  Gray, Gront, Guffy, Horowitz, Huang, Huber, Jacobs, Jeliazkov, Johnson,
  Kappel, Karanicolas, Khakzad, Khar, Khare, Khatib, Khramushin, King,
  Kleffner, Koepnick, Kortemme, Kuenze, Kuhlman, Kuroda, Labonte, Lai,
  Lapidoth, Leaver-Fay, Lindert, Linsky, London, Lubin, Lyskov, Maguire,
  Malmstr{\"o}m, Marcos, Marcu, Marze, Meiler, Moretti, Mulligan, Nerli, Norn,
  {\'O}'Conch{\'u}ir, Ollikainen, Ovchinnikov, Pacella, Pan, Park, Pavlovicz,
  Pethe, Pierce, Pilla, Raveh, Renfrew, Burman, Rubenstein, Sauer, Scheck,
  Schief, Schueler-Furman, Sedan, Sevy, Sgourakis, Shi, Siegel, Silva, Smith,
  Song, Stein, Szegedy, Teets, Thyme, Wang, Watkins, Zimmerman, and
  Bonneau]{rosetta}
Julia~Koehler Leman, Brian~D. Weitzner, Steven~M. Lewis, Jared Adolf-Bryfogle,
  Nawsad Alam, Rebecca~F. Alford, Melanie Aprahamian, David Baker, Kyle~A.
  Barlow, Patrick Barth, Benjamin Basanta, Brian~J. Bender, Kristin Blacklock,
  Jaume Bonet, Scott~E. Boyken, Phil Bradley, Chris Bystroff, Patrick Conway,
  Seth Cooper, Bruno~E. Correia, Brian Coventry, Rhiju Das, Ren{\'e}M. De~Jong,
  Frank DiMaio, Lorna Dsilva, Roland Dunbrack, Alexander~S. Ford, Brandon
  Frenz, Darwin~Y. Fu, Caleb Geniesse, Lukasz Goldschmidt, Ragul Gowthaman,
  Jeffrey~J. Gray, Dominik Gront, Sharon Guffy, Scott Horowitz, Po-Ssu Huang,
  Thomas Huber, Tim~M. Jacobs, Jeliazko~R. Jeliazkov, David~K. Johnson, Kalli
  Kappel, John Karanicolas, Hamed Khakzad, Karen~R. Khar, Sagar~D. Khare, Firas
  Khatib, Alisa Khramushin, Indigo~C. King, Robert Kleffner, Brian Koepnick,
  Tanja Kortemme, Georg Kuenze, Brian Kuhlman, Daisuke Kuroda, Jason~W.
  Labonte, Jason~K. Lai, Gideon Lapidoth, Andrew Leaver-Fay, Steffen Lindert,
  Thomas Linsky, Nir London, Joseph~H. Lubin, Sergey Lyskov, Jack Maguire, Lars
  Malmstr{\"o}m, Enrique Marcos, Orly Marcu, Nicholas~A. Marze, Jens Meiler,
  Rocco Moretti, Vikram~Khipple Mulligan, Santrupti Nerli, Christoffer Norn,
  Shane {\'O}'Conch{\'u}ir, Noah Ollikainen, Sergey Ovchinnikov, Michael~S.
  Pacella, Xingjie Pan, Hahnbeom Park, Ryan~E. Pavlovicz, Manasi Pethe,
  Brian~G. Pierce, Kala~Bharath Pilla, Barak Raveh, P.~Douglas Renfrew, Shourya
  S.~Roy Burman, Aliza Rubenstein, Marion~F. Sauer, Andreas Scheck, William
  Schief, Ora Schueler-Furman, Yuval Sedan, Alexander~M. Sevy, Nikolaos~G.
  Sgourakis, Lei Shi, Justin~B. Siegel, Daniel-Adriano Silva, Shannon Smith,
  Yifan Song, Amelie Stein, Maria Szegedy, Frank~D. Teets, Summer~B. Thyme, Ray
  Yu-Ruei Wang, Andrew Watkins, Lior Zimmerman, and Richard Bonneau.
\newblock {Macromolecular modeling and design in Rosetta: recent methods and
  frameworks}.
\newblock \emph{Nature Methods}, 17\penalty0 (7):\penalty0 665--680, 2020.

\bibitem[Wang et~al.(2008)Wang, Canutescu, and Dunbrack]{scwrl}
Qiang Wang, Adrian~A Canutescu, and Roland~L Dunbrack.
\newblock {SCWRL and MolIDE: Computer programs for side-chain conformation
  prediction and homology modeling}.
\newblock \emph{Nature Protocols}, 3\penalty0 (12):\penalty0 1832--1847, 2008.

\bibitem[Lee and Kim(2025)]{Lee2024}
Jin~Sub Lee and Philip~M. Kim.
\newblock Flowpacker: Protein side-chain packing with torsional flow matching.
\newblock \emph{Bioinformatics}, 41\penalty0 (3):\penalty0 btaf010, 2025.

\bibitem[McPartlon and Xu(2023)]{mcparlton2023}
Matthew McPartlon and Jinbo Xu.
\newblock An end-to-end deep learning method for protein side-chain packing and
  inverse folding.
\newblock \emph{Proceedings of the National Academy of Sciences of the United
  States of America}, 120\penalty0 (23):\penalty0 e2216438120, 2023.

\bibitem[Misiura et~al.(2022)Misiura, Shroff, Thyer, and
  Kolomeisky]{Misiura2022}
Mikita Misiura, Raghav Shroff, Ross Thyer, and Anatoly~B. Kolomeisky.
\newblock {DLPacker: Deep learning for prediction of amino acid side chain
  conformations in proteins}.
\newblock \emph{Proteins: Structure, Function, and Bioinformatics}, 90\penalty0
  (6):\penalty0 1278--1290, 2022.

\bibitem[Visani et~al.(2023)Visani, Galvin, Pun, and Nourmohammad]{visani2023}
Gian~Marco Visani, William Galvin, Michael~Neal Pun, and Armita Nourmohammad.
\newblock {H-Packer}: Holographic rotationally equivariant convolutional neural
  network for protein side-chain packing.
\newblock \emph{arXiv preprint arXiv:2311.09312}, 2023.

\bibitem[Randolph and Kuhlman(2024)]{Randolph2023}
Nicholas~Z. Randolph and Brian Kuhlman.
\newblock Invariant point message passing for protein side chain packing.
\newblock \emph{Proteins: Structure, Function, and Bioinformatics}, 92\penalty0
  (10):\penalty0 1220--1233, 2024.

\bibitem[Jones et~al.(2025)Jones, Khanna, and Ferguson]{jones2025flowback}
Michael~S. Jones, Smayan Khanna, and Andrew~L. Ferguson.
\newblock {FlowBack}: A generalized flow-matching approach for biomolecular
  backmapping.
\newblock \emph{Journal of Chemical Information and Modeling}, 65\penalty0
  (2):\penalty0 672--692, 2025.

\bibitem[Noid et~al.(2008)Noid, Chu, Ayton, Krishna, Izvekov, Voth, Das, and
  Andersen]{noid2008multiscale}
William~George Noid, Jhih-Wei Chu, Gary~S Ayton, Vinod Krishna, Sergei Izvekov,
  Gregory~A Voth, Avisek Das, and Hans~C Andersen.
\newblock {The multiscale coarse-graining method. I. A rigorous bridge between
  atomistic and coarse-grained models}.
\newblock \emph{Journal of Chemical Physics}, 128\penalty0 (24):\penalty0
  244114, 2008.

\bibitem[Clementi(2008)]{clementi2008coarse}
Cecilia Clementi.
\newblock Coarse-grained models of protein folding: toy models or predictive
  tools?
\newblock \emph{Current Opinion in Structural Biology}, 18\penalty0
  (1):\penalty0 10--15, 2008.

\bibitem[Jin et~al.(2022)Jin, Pak, Durumeric, Loose, and Voth]{jin2022bottom}
Jaehyeok Jin, Alexander~J Pak, Aleksander~EP Durumeric, Timothy~D Loose, and
  Gregory~A Voth.
\newblock Bottom-up coarse-graining: Principles and perspectives.
\newblock \emph{Journal of Chemical Theory and Computation}, 18\penalty0
  (10):\penalty0 5759--5791, 2022.

\bibitem[Wassenaar et~al.(2014)Wassenaar, Pluhackova, B{\"o}ckmann, Marrink,
  and Tieleman]{wassenaar2014going}
Tsjerk~A Wassenaar, Kristyna Pluhackova, Rainer~A B{\"o}ckmann, Siewert~J
  Marrink, and D~Peter Tieleman.
\newblock {Going backward: a flexible geometric approach to reverse
  transformation from coarse grained to atomistic models}.
\newblock \emph{Journal of Chemical Theory and Computation}, 10\penalty0
  (2):\penalty0 676--690, 2014.

\bibitem[Lombardi et~al.(2016)Lombardi, Mart{\'\i}, and
  Capece]{lombardi2016cg2aa}
Leandro~E Lombardi, Marcelo~A Mart{\'\i}, and Luciana Capece.
\newblock {CG2AA: Backmapping protein coarse-grained structures}.
\newblock \emph{Bioinformatics}, 32\penalty0 (8):\penalty0 1235--1237, 2016.

\bibitem[Wang et~al.(2022)Wang, Xu, Cai, Miller, Smidt, Wang, Tang, and
  G\'omez-Bombarelli]{wang2022generative}
Wujie Wang, Minkai Xu, Chen Cai, Benjamin~Kurt Miller, Tess Smidt, Yusu Wang,
  Jian Tang, and Rafael G\'omez-Bombarelli.
\newblock Generative coarse-graining of molecular conformations.
\newblock \emph{arXiv preprint arXiv:2201.12176}, 2022.

\bibitem[Jones et~al.(2023)Jones, Shmilovich, and
  Ferguson]{jones2023diamondback}
Michael~S Jones, Kirill Shmilovich, and Andrew~L Ferguson.
\newblock {DiAMoNDBack}: Diffusion-denoising autoregressive model for
  non-deterministic backmapping of {C$\alpha$} protein traces.
\newblock \emph{Journal of Chemical Theory and Computation}, 19\penalty0
  (21):\penalty0 7908--7923, 2023.

\bibitem[Stieffenhofer et~al.(2020)Stieffenhofer, Wand, and
  Bereau]{stieffenhofer2020adversarial}
Marc Stieffenhofer, Michael Wand, and Tristan Bereau.
\newblock Adversarial reverse mapping of equilibrated condensed-phase molecular
  structures.
\newblock \emph{Machine Learning: Science and Technology}, 1\penalty0
  (4):\penalty0 045014, 2020.

\bibitem[Stieffenhofer et~al.(2021)Stieffenhofer, Bereau, and
  Wand]{stieffenhofer2021adversarial}
Marc Stieffenhofer, Tristan Bereau, and Michael Wand.
\newblock Adversarial reverse mapping of condensed-phase molecular structures:
  Chemical transferability.
\newblock \emph{APL Materials}, 9\penalty0 (3):\penalty0 031107, 2021.

\bibitem[Stieffenhofer et~al.(2022)Stieffenhofer, Scherer, May, Bereau, and
  Andrienko]{stieffenhofer2022benchmarking}
Marc Stieffenhofer, Christoph Scherer, Falk May, Tristan Bereau, and Denis
  Andrienko.
\newblock Benchmarking coarse-grained models of organic semiconductors via deep
  backmapping.
\newblock \emph{Frontiers in Chemistry}, 10:\penalty0 982757, 2022.

\bibitem[Li et~al.(2020)Li, Burkhart, Poli\'nska, Harmandaris, and
  Doxastakis]{li2020backmapping}
Wei Li, Craig Burkhart, Patrycja Poli\'nska, Vagelis Harmandaris, and Manolis
  Doxastakis.
\newblock Backmapping coarse-grained macromolecules: An efficient and versatile
  machine learning approach.
\newblock \emph{Journal of Chemical Physics}, 153\penalty0 (4):\penalty0
  041101, 2020.

\bibitem[An and Deshmukh(2020)]{an2020machine}
Yaxin An and Sanket~A Deshmukh.
\newblock Machine learning approach for accurate backmapping of coarse-grained
  models to all-atom models.
\newblock \emph{Chemical Communications}, 56\penalty0 (65):\penalty0
  9312--9315, 2020.

\bibitem[Shmilovich et~al.(2022)Shmilovich, Stieffenhofer, Charron, and
  Hoffmann]{shmilovich2022temporally}
Kirill Shmilovich, Marc Stieffenhofer, Nicholas~E Charron, and Moritz Hoffmann.
\newblock Temporally coherent backmapping of molecular trajectories from
  coarse-grained to atomistic resolution.
\newblock \emph{Journal of Physical Chemistry A}, 126\penalty0 (48):\penalty0
  9124--9139, 2022.

\bibitem[Liu et~al.(2023)Liu, Chen, and Lin]{liu2023backdiff}
Yikai Liu, Ming Chen, and Guang Lin.
\newblock Backdiff: A diffusion model for generalized transferable protein
  backmapping.
\newblock \emph{arXiv preprint arXiv:2310.01768}, 2023.

\bibitem[Li et~al.(2024)Li, Meng, and Liang]{li2024towards}
Jiasheng Li, Zaiqiao Meng, and Shangsong Liang.
\newblock Towards deep generative backmapping of coarse-grained molecular
  systems.
\newblock In \emph{Proceedings of the 2024 2nd Asia Conference on Computer
  Vision, Image Processing and Pattern Recognition}, pages 1--7, 2024.

\bibitem[Pang et~al.(2024)Pang, Yang, and Gumbart]{pang2024simple}
Yui~Tik Pang, Lixinhao Yang, and James~C Gumbart.
\newblock From simple to complex: Reconstructing all-atom structures from
  coarse-grained models using cg2all.
\newblock \emph{Structure}, 32\penalty0 (1):\penalty0 5--7, 2024.

\bibitem[Heo and Feig(2024)]{heo2024one}
Lim Heo and Michael Feig.
\newblock One bead per residue can describe all-atom protein structures.
\newblock \emph{Structure}, 32\penalty0 (1):\penalty0 97--111, 2024.

\bibitem[Angioletti et~al.(2024)Angioletti, Raniolo, and
  Limongelli]{angioletti2024herobm}
Daniele Angioletti, Stefano Raniolo, and Vittorio Limongelli.
\newblock {HEroBM}: A deep equivariant graph neural network for universal
  backmapping from coarse-grained to all-atom representations.
\newblock \emph{arXiv preprint arXiv:2404.16911}, 2024.

\bibitem[Waltmann et~al.(2025)Waltmann, Wang, Yang, Kim, and
  Voth]{waltmann2025msback}
Curt Waltmann, Yihang Wang, Chengxi Yang, Siyoung Kim, and Gregory~A Voth.
\newblock {MSBack}: Multiscale backmapping of highly coarse-grained proteins
  using constrained diffusion.
\newblock \emph{Journal of Chemical Theory and Computation}, 2025.

\bibitem[Berman et~al.(2000)Berman, Westbrook, Feng, Gilliland, Bhat, Weissig,
  Shindyalov, and Bourne]{10.1093/nar/28.1.235}
Helen~M. Berman, John Westbrook, Zukang Feng, Gary Gilliland, T.~N. Bhat, Helge
  Weissig, Ilya~N. Shindyalov, and Philip~E. Bourne.
\newblock {The Protein Data Bank}.
\newblock \emph{Nucleic Acids Research}, 28\penalty0 (1):\penalty0 235--242,
  2000.

\bibitem[Domingo-Enrich et~al.(2025)Domingo-Enrich, Drozdzal, Karrer, and
  Chen]{domingoenrich2025adjointmatching}
Carles Domingo-Enrich, Michal Drozdzal, Brian Karrer, and Ricky T.~Q. Chen.
\newblock Adjoint matching: Fine-tuning flow and diffusion generative models
  with memoryless stochastic optimal control.
\newblock \emph{arXiv preprint arXiv:2409.08861}, 2025.

\bibitem[Souza et~al.(2021)Souza, Alessandri, Barnoud, Thallmair, Faustino,
  Gr{\"u}newald, Patmanidis, Abdizadeh, Bruininks, Wassenaar, Kroon, Melcr,
  Nieto, Corradi, Khan, Domanski, Javanainen, Martinez-Seara, Reuter, Best,
  Vattulainen, Monticelli, Periole, Tieleman, de~Vries, and
  Marrink]{souza2021martini}
Paulo C.~T. Souza, Riccardo Alessandri, Jonathan Barnoud, Sebastian Thallmair,
  Ignacio Faustino, Fabian Gr{\"u}newald, Ilias Patmanidis, Haleh Abdizadeh,
  Bart M.~H. Bruininks, Tsjerk~A. Wassenaar, Peter~C. Kroon, Josef Melcr,
  Vincent Nieto, Valentina Corradi, Hanif~M. Khan, Jan~J. Domanski, Matti
  Javanainen, Hector Martinez-Seara, Nathalie Reuter, Robert~B. Best, Ilpo
  Vattulainen, Luca Monticelli, Xavier Periole, D.~Peter Tieleman, Alex~H
  de~Vries, and Siewert~J. Marrink.
\newblock Martini 3: A general purpose force field for coarse-grained molecular
  dynamics.
\newblock \emph{Nature Methods}, 18:\penalty0 382--388, 2021.

\bibitem[Davtyan et~al.(2012)Davtyan, Schafer, Zheng, Clementi, Wolynes, and
  Papoian]{davtyan2012awsem}
Aram Davtyan, Nicholas~P Schafer, Weihua Zheng, Cecilia Clementi, Peter~G
  Wolynes, and Garegin~A Papoian.
\newblock {AWSEM-MD: Protein structure prediction using coarse-grained physical
  potentials and bioinformatically based local structure biasing}.
\newblock \emph{Journal of Physical Chemistry B}, 116\penalty0 (29):\penalty0
  8494--8503, 2012.

\bibitem[Tesei and Lindorff-Larsen(2023)]{tesei2023improved}
Giulio Tesei and Kresten Lindorff-Larsen.
\newblock Improved predictions of phase behaviour of intrinsically disordered
  proteins by tuning the interaction range.
\newblock \emph{Open Research Europe}, 2:\penalty0 94, 2023.

\bibitem[Geffner et~al.(2025)Geffner, Didi, Cao, Reidenbach, Zhang, Dallago,
  Kucukbenli, Kreis, and Vahdat]{geffner2025laproteina}
Tomas Geffner, Kieran Didi, Zhonglin Cao, Danny Reidenbach, Zuobai Zhang,
  Christian Dallago, Emine Kucukbenli, Karsten Kreis, and Arash Vahdat.
\newblock {La-Proteina}: Atomistic protein generation via partially latent flow
  matching.
\newblock \emph{arXiv preprint arXiv:2507.09466}, 2025.

\bibitem[Qu et~al.(2025)Qu, Guan, Ma, Zhai, Wu, and Wang]{Qu2024}
Wei Qu, Jiawei Guan, Rui Ma, Ke~Zhai, Weikun Wu, and Haobo Wang.
\newblock P(all-atom) is unlocking new path for protein design.
\newblock \emph{bioRxiv preprint bioRxiv:10.1101/2024.08.16.608235}, 2025.

\bibitem[Lu et~al.(2025)Lu, Chen, Lu, Lozano, Chenthamarakshan, Das, and
  Tang]{lu2025eba}
Jiarui Lu, Xiaoyin Chen, Stephen~Zhewen Lu, Aur{\'e}lie Lozano, Vijil
  Chenthamarakshan, Payel Das, and Jian Tang.
\newblock Aligning protein conformation ensemble generation with physical
  feedback.
\newblock \emph{arXiv preprint arXiv:2505.24203}, 2025.

\bibitem[King and Koes(2021)]{king2021sidechainnet}
Jonathan~Edward King and David~Ryan Koes.
\newblock {SidechainNet}: An all-atom protein structure dataset for machine
  learning.
\newblock \emph{Proteins: Structure, Function, and Bioinformatics}, 89\penalty0
  (11):\penalty0 1489--1496, 2021.

\bibitem[AlQuraishi(2019)]{alquraishi2019proteinnet}
Mohammed AlQuraishi.
\newblock {ProteinNet}: a standardized data set for machine learning of protein
  structure.
\newblock \emph{BMC Bioinformatics}, 20\penalty0 (1):\penalty0 1--10, 2019.

\bibitem[MacKerell~Jr. et~al.(2000)MacKerell~Jr., Banavali, and
  Foloppe]{CHARMM}
Alexander~D. MacKerell~Jr., Nilesh Banavali, and Nicolas Foloppe.
\newblock Development and current status of the {CHARMM} force field for
  nucleic acids.
\newblock \emph{Biopolymers}, 56\penalty0 (4):\penalty0 257--265, 2000.

\bibitem[Dill(1990)]{dill1990dominant}
Ken~A Dill.
\newblock Dominant forces in protein folding.
\newblock \emph{Biochemistry}, 29\penalty0 (31):\penalty0 7133--7155, 1990.

\bibitem[Aldeghi et~al.(2019)Aldeghi, Gapsys, and de~Groot]{aldeghi2019}
Matteo Aldeghi, Vytautas Gapsys, and Bert~L. de~Groot.
\newblock Predicting kinase inhibitor resistance: Physics-based and data-driven
  approaches.
\newblock \emph{ACS Central Science}, 5\penalty0 (8):\penalty0 1468--1474,
  2019.

\bibitem[Domingo-Enrich et~al.(2024)Domingo-Enrich, Han, Amos, Bruna, and
  Chen]{domingoenrich2024socmatching}
Carles Domingo-Enrich, Jiequn Han, Brandon Amos, Joan Bruna, and Ricky T.~Q.
  Chen.
\newblock Stochastic optimal control matching.
\newblock \emph{arXiv preprint arXiv:2312.02027}, 2024.

\bibitem[Ma et~al.(2024)Ma, Goldstein, Albergo, Boffi, Vanden-Eijnden, and
  Xie]{ma2024}
Nanye Ma, Mark Goldstein, Michael~S. Albergo, Nicholas~M. Boffi, Eric
  Vanden-Eijnden, and Saining Xie.
\newblock {SiT:} exploring flow and diffusion-based generative models with
  scalable interpolant transformers.
\newblock \emph{arXiv preprint arXiv:2401.08740}, 2024.

\bibitem[Kappen(2005)]{Kappen_2005}
H~J Kappen.
\newblock Path integrals and symmetry breaking for optimal control theory.
\newblock \emph{Journal of Statistical Mechanics: Theory and Experiment},
  2005\penalty0 (11):\penalty0 P11011, 2005.

\bibitem[Kingma and Ba(2014)]{kingma2014adam}
Diederik~P Kingma and Jimmy Ba.
\newblock Adam: A method for stochastic optimization.
\newblock \emph{arXiv preprint arXiv:1412.6980}, 2014.

\bibitem[Lindorff-Larsen et~al.(2011)Lindorff-Larsen, Piana, Dror, and
  Shaw]{DESRESTrajs}
Kresten Lindorff-Larsen, Stefano Piana, Ron~O. Dror, and David~E. Shaw.
\newblock How fast-folding proteins fold.
\newblock \emph{Science}, 334\penalty0 (6055):\penalty0 517--520, 2011.

\bibitem[Abraham et~al.(2015)Abraham, Murtola, Schulz, P{\'a}ll, Smith, Hess,
  and Lindahl]{ABRAHAM201519}
Mark~James Abraham, Teemu Murtola, Roland Schulz, Szil{\'a}rd P{\'a}ll,
  Jeremy~C. Smith, Berk Hess, and Erik Lindahl.
\newblock {GROMACS}: {High} performance molecular simulations through
  multi-level parallelism from laptops to supercomputers.
\newblock \emph{SoftwareX}, 1-2:\penalty0 19--25, 2015.

\bibitem[Eastman et~al.(2017)Eastman, Swails, Chodera, McGibbon, Zhao,
  Beauchamp, Wang, Simmonett, Harrigan, Stern, Wiewiora, Brooks, and
  Pande]{Eastman2017OpenMM7}
Peter Eastman, Jason Swails, John~D. Chodera, Robert~T. McGibbon, Yonghai Zhao,
  Kyle~A. Beauchamp, Lee-Ping Wang, Andrew~C. Simmonett, Matthew~P. Harrigan,
  Chaya~D. Stern, Rafal~P. Wiewiora, Bernard~R. Brooks, and Vijay~S. Pande.
\newblock {OpenMM} 7: Rapid development of high performance algorithms for
  molecular dynamics.
\newblock \emph{PLOS Computational Biology}, 13\penalty0 (7):\penalty0
  e1005659, 2017.

\bibitem[Darden et~al.(1993)Darden, York, and Pedersen]{PME}
Tom Darden, Darrin York, and Lee Pedersen.
\newblock {Particle mesh Ewald: An N log(N) method for Ewald sums in large
  systems}.
\newblock \emph{The Journal of Chemical Physics}, 98\penalty0 (12):\penalty0
  10089--10092, 06 1993.
\newblock ISSN 0021-9606.

\bibitem[Stillinger(1999)]{stillinger1999exponential}
Frank~H Stillinger.
\newblock Exponential multiplicity of inherent structures.
\newblock \emph{Physical Review E}, 59\penalty0 (1):\penalty0 48, 1999.

\bibitem[Kitts et~al.(2019)Kitts, Giglio, Zamorano-S{\'a}nchez, Park, Townsley,
  Cooley, Wucher, Klose, Nadell, Yildiz, and Sondermann]{Kitts2019}
Giordan Kitts, Krista~M. Giglio, David Zamorano-S{\'a}nchez, Jin~Hwan Park,
  Loni Townsley, Richard~B. Cooley, Benjamin~R. Wucher, Karl~E. Klose, Carey~D.
  Nadell, Fitnat~H. Yildiz, and Holger Sondermann.
\newblock A conserved regulatory circuit controls large adhesins in vibrio
  cholerae.
\newblock \emph{mBio}, 10\penalty0 (6), 2019.

\bibitem[Johansson et~al.(1997)Johansson, {de Ch{\^a}teau}, Wikstr{\"o}m,
  Fors{\'e}n, Drakenberg, and Bj{\"o}rck]{JOHANSSON1997859}
Maria~U Johansson, Maarten {de Ch{\^a}teau}, Mats Wikstr{\"o}m, Sture
  Fors{\'e}n, Torbj{\"o}rn Drakenberg, and Lars Bj{\"o}rck.
\newblock {Solution structure of the albumin-binding GA module: A versatile
  bacterial protein domain}.
\newblock \emph{Journal of Molecular Biology}, 266\penalty0 (5):\penalty0
  859--865, 1997.

\bibitem[Pace et~al.(2011)Pace, Fu, Fryar, Landua, Trevino, Shirley, Hendricks,
  Iimura, Gajiwala, Scholtz, and Grimsley]{Pace2011-vh}
C.~Nick Pace, Hailong Fu, Katrina~Lee Fryar, John Landua, Saul~R Trevino,
  Bret~A Shirley, Marsha~Mcnutt Hendricks, Satoshi Iimura, Ketan Gajiwala,
  J~Martin Scholtz, and Gerald~R Grimsley.
\newblock Contribution of hydrophobic interactions to protein stability.
\newblock \emph{Journal of Molecular Biology}, 408\penalty0 (3):\penalty0
  514--528, 2011.

\bibitem[Geney et~al.(2006)Geney, Layten, Gomperts, Hornak, and
  Simmerling]{Geney2006}
{Rapha{\"e}l} Geney, Melinda Layten, Roberto Gomperts, Viktor Hornak, and
  Carlos Simmerling.
\newblock Investigation of salt bridge stability in a generalized born solvent
  model.
\newblock \emph{Journal of Chemical Theory and Computation}, 2\penalty0
  (1):\penalty0 115--127, 2006.

\bibitem[Wood et~al.(2025)Wood, Dzamba, Fu, Gao, Shuaibi, Barroso-Luque,
  Abdelmaqsoud, Gharakhanyan, Kitchin, Levine, Michel, Sriram, Cohen, Das,
  Rizvi, Sahoo, Ulissi, and Zitnick]{wood2025uma}
Brandon~M. Wood, Misko Dzamba, Xiang Fu, Meng Gao, Muhammed Shuaibi, Luis
  Barroso-Luque, Kareem Abdelmaqsoud, Vahe Gharakhanyan, John~R. Kitchin,
  Daniel~S. Levine, Kyle Michel, Anuroop Sriram, Taco Cohen, Abhishek Das,
  Ammar Rizvi, Sushree~Jagriti Sahoo, Zachary~W. Ulissi, and C.~Lawrence
  Zitnick.
\newblock {UMA}: A family of universal models for atoms.
\newblock \emph{arXiv preprint arXiv:2506.23971}, 2025.

\bibitem[Havens et~al.(2025)Havens, Miller, Yan, Domingo-Enrich, Sriram, Wood,
  Levine, Hu, Amos, Karrer, Fu, Liu, and Chen]{havens2025}
Aaron Havens, Benjamin~Kurt Miller, Bing Yan, Carles Domingo-Enrich, Anuroop
  Sriram, Brandon Wood, Daniel Levine, Bin Hu, Brandon Amos, Brian Karrer,
  Xiang Fu, Guan-Horng Liu, and Ricky T.~Q. Chen.
\newblock Adjoint sampling: Highly scalable diffusion samplers via adjoint
  matching.
\newblock \emph{arXiv preprint arXiv:2504.11713}, 2025.

\end{thebibliography}
\bibliographystyle{unsrtnat}

\end{document}